\documentstyle[11pt,aaspp4,tighten,flushrt]{article}

\newcommand{\kms}{\mbox{${\rm\,km\,s}^{-1}$}}

\newcommand{\pc}{\mbox{$\rm\,pc$}}
\newcommand{\kpc}{\mbox{$\rm\,kpc$}}

\newcommand{\msun}{\mbox{$M_\odot$}}
\newcommand{\microm}{\mbox{$\rm\,\mu m$}}

\newcommand{\ang}{\mbox{$\rm\,\AA$}}
\newcommand{\beq}{\begin{equation}}
\newcommand{\eeq}{\end{equation}}

\lefthead{Statler \& Smecker-Hane}
\righthead{NGC 3379}

\begin{document}

\title{The Stellar Kinematic Fields of NGC 3379\footnotemark}

\footnotetext{Observations reported in this paper were obtained at the
Multiple Mirror Telescope Observatory, a joint facility of the University
of Arizona and the Smithsonian Institution.}

\author{Thomas S. Statler}
\affil{Department of Physics and Astronomy, Ohio University, Athens, OH 45701,
USA; tss@coma.phy.ohiou.edu}
\and
\author{Tammy Smecker-Hane}
\affil{Department of Physics, University of California, Irvine, CA 92717, USA;
smecker@carina.ps.uci.edu}

\vskip -3.5in {\hfill \sl To appear in The Astronomical Journal, 1999 February}
\vskip 3.4in

\begin{abstract}

We have measured the stellar kinematic profiles of NGC 3379 along four
position angles, using absorption lines in spectra obtained with the Multiple
Mirror Telescope. We derive a far more detailed description of the kinematic
fields through the main body of the galaxy than could be obtained from
previous work. Our data extend $90\arcsec$ from the center, at
essentially seeing-limited resolution out to $17\arcsec$. The derived
mean velocities and dispersions have total errors (internal and systematic)
better than $\pm 10\kms$, and frequently better than $5\kms$, out to
$55\arcsec$. We find very weak ($3\kms$) rotation on the minor axis
interior to $12\arcsec$, and no detectable rotation above $6\kms$ from
$12\arcsec$ to $50\arcsec$ or above $16\kms$ out to $90\arcsec$ (95\% confidence
limits). However, a Fourier reconstruction of the mean velocity field from
all 4 sampled PAs does indicate a $\sim 5\arcdeg$ twist of the kinematic major
axis, in the direction opposite to the known isophotal twist. The $h_3$ and
$h_4$ parameters are found to be generally small over the entire observed
region. The azimuthally-averaged dispersion profile joins smoothly at large
radii with the velocity dispersions of planetary nebulae.
Unexpectedly, we find bends in the major-axis rotation
curve, also visible (though less pronounced) on the diagonal position angles.
The outermost bend closely coincides in position with other sharp
kinematic features: an abrupt flattening of the dispersion profile,
and local peaks in $h_3$ and $h_4$. All of these features are in a
photometrically interesting region where the surface brightness profile
departs significantly from an $r^{1/4}$ law. Features such as these are
not generally known in ellipticals owing to a lack of data at comparable
resolution. Very similar behavior, however, is seen the kinematics of
the edge-on S0 NGC 3115. We discuss the suggestion that NGC 3379 could
be a misclassified S0; preliminary results from dynamical modeling indicate
that it may be a flattened, weakly triaxial system seen in an orientation
that makes it appear round.

\end{abstract}


\section{Introduction}

After M87 and Centaurus A, NGC 3379 (M105) is one of the best-studied
elliptical galaxies in the sky. Virtually a walking advertisement for
the $r^{1/4}$ law (de Vaucouleurs \& Capaccioli \markcite{dVC79}1979),
this system is regularly used as a control object or calibrator for a
variety of photometric and spectroscopic studies. NGC 3379 has all the
hallmarks of a ``classic'' early-type galaxy: almost perfectly elliptical
isophotes and colors characteristic of an old stellar population (Peletier
et al.\ \markcite{Pel90}1990, Goudfrooij et al.\ \markcite{Gou94}1994);
slow rotation about the apparent minor axis (Davies \& Illingworth
\markcite{DaI83}1983, Davies \& Birkinshaw \markcite{DaB88}1988);
no shells, tails, or other signs of interactions (Schweizer \& Seitzer
\markcite{ScS92}1992); no detection in either \ion{H}{1} (Bregman, Hogg,
\& Roberts \markcite{BHR92}1992) or CO (Sofue \& Wakamatsu
\markcite{SoW93}1993); very modest H$\alpha$+[\ion{N}{2}] emission
(Macchetto et al.\ \markcite{Mac96}1996); and only minimal absorption by
dust in the inner $4\arcsec$ (van Dokkum \& Franx \markcite{vDF95}1995,
Michard \markcite{Mic98}1998).

Yet, for all its familiarity, there are serious questions as to the
true nature of our ``standard elliptical.'' For one, there is a nagging
concern that it might not be an elliptical at all. Capaccioli and
collaborators (Capaccioli \markcite{Cap87}1987; Nieto, Capaccioli, \& Held 
\markcite{NCH88}1988; Capaccioli, Held, Lorenz, \& Vietri
\markcite{Cap90}1990; Capaccioli, Vietri, Held, \& Lorenz
\markcite{Cap91}1991, hereafter CVHL) have argued, mainly on photometric
grounds, that NGC 3379 could be a misclassified S0 seen close to face-on.
CVHL demonstrate that a deprojected spheroid+disk model for the edge-on
S0 NGC 3115, seen face-on, would show deviations from the best-fit
$r^{1/4}$ law very similar to the $\sim 0.1$ magnitude ripple-like
residuals seen in NGC 3379. They propose that NGC 3379
could be a triaxial S0, since a triaxiality gradient could explain
the observed $5\arcdeg$ isophotal twist.

Statler \markcite{Sta94}(1994) has also examined the shape of NGC 3379,
using dynamical models to fit surface photometry
(Peletier et al.\ \markcite{Pel90}1990) and multi-position-angle velocity
data (Davies \& Birkinshaw \markcite{DaB88}1988, Franx,
Illingworth, \& Heckman \markcite{FIH89}1989). The 
data are found to rule out very flattened, highly triaxial shapes
such as that suggested by CVHL, while still being consistent with either
flattened axisymmetric or rounder triaxial figures. The results are limited,
however, by the accuracy of the kinematic data, which are unable to
constrain the rotation on the minor axis beyond $R=15\arcsec$ to any better
than $30\%$ of the peak major-axis velocity. This large an
uncertainty implies a $\sim 30\arcdeg$ ambiguity in the position of the
apparent rotation axis. Moreover, there are hints of steeply increasing
minor-axis rotation beyond $30\arcsec$. It is far from clear from the
current data that the common perception of NGC 3379 as a
``classic major-axis rotator'' is an accurate description of the galaxy
beyond---or even at---one effective radius.

Deeper, higher accuracy spectroscopic data are needed, both to define more
precisely the kinematic structure of the galaxy at moderate radii, and also to
establish the connection with the large-$R$ kinematics as determined from
planetary nebulae. Ciardullo et al.\ \markcite{CJD93}(1993) find that
the velocity dispersion in the PN population declines steadily with radius,
reaching $\sim 70\kms$ at $170\arcsec$ (roughly 3 effective radii). This
decline is consistent with a Keplerian falloff outside $1 r_e$, apparently
making NGC 3379 one of the strongest cases for an elliptical galaxy
with a constant mass-to-light ratio and no significant contribution
from dark matter inside $9\kpc$. On the other hand, if the PNe were in a
nearly face-on disk, the line-of-sight dispersion may not reflect
the true dynamical support of the system. To correctly interpret the PN
data, therefore, one needs to know how the stellar data join onto
the PN dispersion profile, as well as have a good model for the shape and
orientation of the galaxy.

At small $R$, {\em HST\/} imaging shows NGC 3379 to be a ``core galaxy'';
i.e., its surface brightness profile turns over near $1\arcsec$ --
$2\arcsec$ to an inner logarithmic slope of about $-0.18$ (Byun et al.\
\markcite{Byu96}1996). A non-parametric deprojection assuming spherical
symmetry (Gebhardt et al.\ \markcite{Geb96}1996) gives a logarithmic
slope in the {\em volume\/} luminosity density of $-1.07 \pm 0.06$
at $r=0\farcs 1$ ($5\pc$). This is rather a shallow slope for galaxies
of this luminosity ($M_V=-20.55$), and is actually more characteristic of
galaxies some 4 times as luminous (Gebhardt et al.\ \markcite{Geb96}1996).
At the same time, NGC 3379 is
a likely candidate for harboring a central dark mass of several hundred
million $\msun$ (Magorrian et al.\ \markcite{Mag98}1998). Since both
density cusps and central point masses have been implicated as potential
saboteurs of triaxiality through orbital chaos (Merritt \& Fridman
\markcite{MeF96}1996, Merritt \& Valluri \markcite{MeV96}1996, Merritt
\markcite{Mer97}1997), a measurement of triaxiality from the stellar
kinematics would be valuable in gauging the importance of this mechanism
in real systems.

Here we present new spectroscopic observations of NGC 3379, as part
of our program to obtain multi-position-angle kinematic data at high
accuracy and good spatial resolution for a sample of photometrically
well-studied ellipticals. We obtain a far more detailed rendition of the
kinematic fields through the main body of the galaxy
than has been available from previous data. We find that
these fields suggest a two-component structure for the galaxy, and closely
resemble those of the S0 NGC 3115. We reserve firm
conclusions on the shape and Hubble type of NGC 3379 for a later paper
devoted to dynamical modeling; here we present the data. Section
\ref{s.observations} of this paper describes the observational
procedure. Data reduction techniques are detailed in Sec.\ \ref{s.reduction},
and the results are presented in Sec.\ \ref{s.results}. We compare our
data with previous work and discuss some of the implications for the structure
of the galaxy in Sec.\ \ref{s.discussion}, and Sec.\ \ref{s.conclusions}
concludes. 

\section{Observations\label{s.observations}}

NGC 3379 was observed with the Multiple Mirror Telescope and the Red Channel
Spectrograph (Schmidt et al.\ \markcite{Sch89}1989) on 3--4 February 1995
UT. The $1\farcs0 \times 180\arcsec$ slit was used with the 1200 grooves/mm
grating to give a resolution of approximately $2.2\ang$ and a spectral
coverage from $\lambda\lambda$ $4480$ -- $5480\ang$. The spectra were imaged
on the $1200\times 800$ Loral CCD ($15 \microm$ pixels, 1 pix = $0\farcs 3$,
read noise = 7 $e^-$), resulting in a nominal dispersion of $0.72 \ang$/pix.
The CCD was read-binned $1 \times 4$ pixels in the dispersion $\times$
spatial directions to reduce read noise, so that the final spatial scale
was $1\farcs2$ per binned pixel.

Except for a brief period of fluctuating seeing on the first night, all
data were taken in photometric conditions. NGC 3379 was observed at four
slit position angles: PA = $70\arcdeg$ (major axis), $340\arcdeg$ (minor
axis), $25\arcdeg$, and $115\arcdeg$. PA 340 was observed entirely on
night 1, PAs 70 and 115 on night 2, and PA 25 over both nights. Four
exposures of 1800 s each were obtained at each PA, except for the last
exposure at PA 70 which was shortened to 900 s due to impending twilight.
Because the galaxy filled the slit, separate 600 s blank sky exposures
were obtained at 30 -- 90 minute intervals depending on the elevation of
the galaxy. Comparison arc spectra were taken before and/or after each
galaxy and sky exposure.

In addition to the standard calibration frames, spectra of radial velocity
standard, flux standard, and Lick/IDS library stars were taken during
twilight. The Lick stars were chosen to have a range of spectral types
and metallicities in order to create composite spectral templates
and to calibrate measurements of line strength indices (to be
presented in a future paper). Stars were trailed across the slit to
illuminate it uniformly. This was an essential step in producing accurate
kinematic profiles because our slit width was wider than the seeing
disk; fits to the spatial profiles of all stellar spectra
give a mean Gaussian width of the point spread function of $0\farcs83$,
with a standard deviation of $0\farcs09$.

\section{Data Reduction\label{s.reduction}}

\subsection{Initial Procedures\label{s.initial}}

Basic reductions were performed as described by Statler, Smecker-Hane, \&
Cecil \markcite{SSC96}(1996, hereafter SSC), using standard
procedures in IRAF. The initial
processing consisted of overscan and bias corrections, flat fielding,
and removal of cosmic rays. This was followed by wavelength calibration
from the comparison arcs, and straightening of all spectra using stellar
traces at different positions along the slit. We used ``unweighted
extraction'' to derive one-dimensional stellar spectra, and rebinned all
spectra onto a logarithmic wavelength scale with pixel width $\Delta x
\equiv \Delta \ln \lambda = 1.626 \times 10^{-4}$ ($\Delta v = 48.756 \kms$).
In the same transformation, the galaxy frames for each PA were registered
spatially.

Time-weighted average sky spectra were created for each galaxy frame by
combining the two sky frames, $Y(t_1)$ and $Y(t_2)$, taken before and
after the galaxy frame $G(t)$. (Times refer to the middle of
the exposures.) The combined sky image was
$Y = K[aY(t_1) + (1-a) Y(t_2)]$, where $a=(t_2-t)/(t_2-t_1)$ and the
constant $K$ ($=3$ for all frames but one) scaled the image to the
exposure time of $G$. Because conditions were photometric, there was
no need to fine-tune $a$ and $K$ by hand, as was done by SSC to improve
the removal of the bright sky emission lines. To avoid degrading the
signal-to-noise ratio in the regions where accurate sky subtraction was
most crucial, the sky spectra were averaged in the spatial direction
by smoothing with a variable-width boxcar window. The width of the window
increased from 1 pixel at the center of the galaxy to 15 pixels at the slit
ends. Finally, after subtracting the smoothed sky images, the 4 galaxy
frames at each PA were coadded.

In parallel with the above procedure, we performed an alternative sky
subtraction in order to estimate the systematic error associated with
this part of the reduction. In the alternative method we simply
subtracted the sky exposure closest in time to each galaxy frame, scaled
up to the appropriate exposure time and boxcar-smoothed. These
``naive sky'' results will be discussed in Sec.\ \ref{s.systematic} below.

SSC worried extensively about the effect of scattered light on the
derived kinematics at large radii. Using their 2-D stellar spectra, they
constructed an approximate smoothed model of the scattered light
contribution and subtracted it from their coadded spectra of NGC 1700.
We have not attempted to do this here, for three reasons. First, we
found that the scattered light characteristics of the Red Channel had
changed significantly from 1993 to 1995, and could no longer be
modeled simply. Second, SSC had noted that the scattered-light correction
resulted in only tiny changes to their kinematic profiles, and that the
contribution to the systematic error budget was negligible
compared to those from sky subtraction and template mismatch. Finally,
NGC 3379 is much less centrally concentrated than NGC
1700, and therefore is much less prone to scattered-light
contamination since the galaxy is still fairly bright even at the ends of
the slit.

The 2-D galaxy spectra were binned into 1-D spectra with the bin width
chosen to preserve a signal-to-noise ratio $\gtrsim 50$ per pixel over most
of the slit length; $S/N$ decreases to around $30$ in the second-to-outermost
bins, and to $20$ in the last bins, which terminate at the end of the slit.
These last bins also suffer a slight degradation in focus, so that the
velocity dispersion
is likely to be overestimated there. The 1-D spectra were divided by
smooth continua fitted using moderate-order cubic splines.
Residual uncleaned cosmic rays and imperfectly subtracted sky lines
were replaced with linear interpolations plus Gaussian noise.
Spectra were tapered over the last 64 pixels at either end
and padded out to a length of 1300 pixels.

The velocity zero point was set using 5 spectra of the IAU radial velocity
standards HD 12029, HD 23169 (observed twice), HD 32963, and HD 114762.
The spectra were shifted to zero velocity and all 10 pairs were
cross-correlated as a consistency check. In only 3 cases were the derived
residual shifts greater than $0.01\kms$ and in no case were they greater
than $0.7\kms$. The velocities of the remaining 20 stars were then found by
averaging the results of cross-correlation against each of the standards,
and these stars were also shifted to zero velocity.

\subsection{LOSVD Extraction\label{s.extraction}}

Parametric extraction of the line-of-sight velocity distributions (LOSVDs)
was performed using
Statler's \markcite{Sta95}(1995) implementation of the cross-correlation
(XC) method, which follows from the relationship between the galaxy
spectrum $G(x)$, the observed spectral template $S(x)$, and the ``ideal
template'' $I(x)$---a zero-velocity composite spectrum of the actual mix of
stars in the galaxy. This relationship is given by
\beq\label{e.crosscorr}
G \circ S = (I \circ S) \otimes B,
\eeq
where $\otimes$ denotes convolution, $\circ$ denotes correlation,
and $B(x)$, the broadening function, is the LOSVD written as a
function of $v/c$. Since the ideal template is unknown, one
replaces $(I \circ S)$ with the template autocorrelation function
$A=S \circ S$, and then manipulates $B$ so that its convolution with $A$
fits the primary peak of the cross-correlation function
$X=G \circ S$. We adopted a Gauss-Hermite expansion for the LOSVD
(van der Marel \& Franx \markcite{vdMF93}1993):
\beq\label{e.losvd}
L(v) = {\gamma \over (2 \pi)^{1/2} \sigma}
\left[1 + h_3 {(2 w^3 - 3 w)\over 3^{1/2}} + h_4
{(4 w^4 - 12 w^2 + 3) \over 24^{1/2}} \right] e^{-w^2/2}, \qquad
w \equiv {v - V \over \sigma}.
\eeq
The expansion was truncated at $h_4$, and non-negativity of $L(v)$ was
enforced by cutting off the tails of the distribution beyond the first
zeros on either side of the center.
Because the XC method can be confused
by broad features in the spectra unrelated to Doppler broadening,
it was necessary to filter out low-frequency components before
cross-correlating. Our adopted filter was zero below a threshold wavenumber
$k_L$ (measured in inverse pixels), unity above $2k_L$, and joined by a
cosine taper in between. More conveniently we can quote the filter width in
Fourier-space pixels as a quantity $W_T=1300k_L$. Empirically we found our
results to be insensitive to $W_T$ over a range centered around
$W_T=15$, which value we adopted for all subsequent analysis.

A non-parametric approach also rooted in equation (\ref{e.crosscorr})
is the Fourier Correlation Quotient method (Bender \markcite{Ben90}1990),
which operates in the Fourier domain. Denoting the Fourier
transform by $\tilde{\ }$, we have
\beq\label{e.fcq}
\tilde{B} = \tilde{X}/\tilde{A};
\eeq
thus $B$ can, in principle, be obtained directly. However, the FCQ method
requires that, to avoid amplifying nose, {\em high\/} frequency components
also be filtered out of the data. This is generally done using an optimal
filter, the construction of which is not an entirely objective procedure
when $S/N \lesssim 50$. We present results from the FCQ method in section
\ref{s.nonparametric}.

\subsection{Composite Templates\label{s.templates}}

Sixteen stars with spectral types between G0 and M1 were available to be
used as templates. We first computed the kinematic profiles for the major
axis (PA 70) using all 16 templates, then set out to choose a set of 4
from which to construct composites. We found, as did SSC, that the
algebraic problem of fitting the galaxy spectrum with a set of very similar
stellar spectra becomes seriously ill-conditioned with more than 4 in
the library. Coefficients were calculated to optimize the fit to the
galaxy spectrum, using a random search of the parameter space as
described by SSC.

For the most part, kinematic profiles derived using different templates
had similar shapes but with different constant offsets, in agreement with
the results of SSC and others (e.g., Rix \& White
\markcite{RiW92}1992, van der Marel et al.\ \markcite{vdM94}1994).
A few templates could be discarded for giving wildly discrepant results.
In principle, a semi-objective criterion for choosing a
library ought to have been available from the requirement that the $h_3$
profile be antisymmetric across the center of the galaxy. However, every
template gave positive values of $h_3$ at $R=0$; we attribute this to
the well-documented discordancy between Mg and Fe line strengths in
ellipticals relative to population-synthesis models with solar Mg/Fe ratio
(Peletier \markcite{Pel89}1989, Gonzalez \markcite{Gon93}1993, Davies
\markcite{Dav96}1996 and references therein).
We therefore proceeded by trial and error, requiring that (1) weight be
distributed roughly evenly among the library spectra in the derived composites;
(2) the central $h_3$ values computed from the composites come out close to
zero; (3) the values of the line strength parameter $\gamma$ come out not
very far from unity; and (4) as wide a range as possible of spectral types and
metallicities be represented.

We found that acceptable composite templates could be constructed at all
positions in the galaxy, consistent with the above criteria, using the
following stars: HD 41636 (G9III), HD 145328 (K0III-IV), HD 132142 (K1V),
and HD 10380 (K3III). We constructed a separate composite at each radius and
position angle. The coefficients of the individual spectra varied, for the most
part, smoothly with radius, from average central values of
$(0.05,0.1,0.4,0.45)$ to roughly $(0.2,0.15,0.4,0.25)$
at large radii. The point-to-point scatter in the coefficients
exceeded $0.1$ outside of about $10\arcsec$ and $0.2$ beyond $30\arcsec$.
However, we saw no indication of this scatter inducing any systematic
effects in the kinematic results beyond those discussed in the next section.

All of the results presented in this paper use the template stars listed above;
however, the analysis was also carried through using an earlier,
unsatisfactory library in order to estimate the systematic error from residual
template mismatch in the composites.

\begin{figure}[t]{\hfill\epsfxsize=3.7in\epsfbox{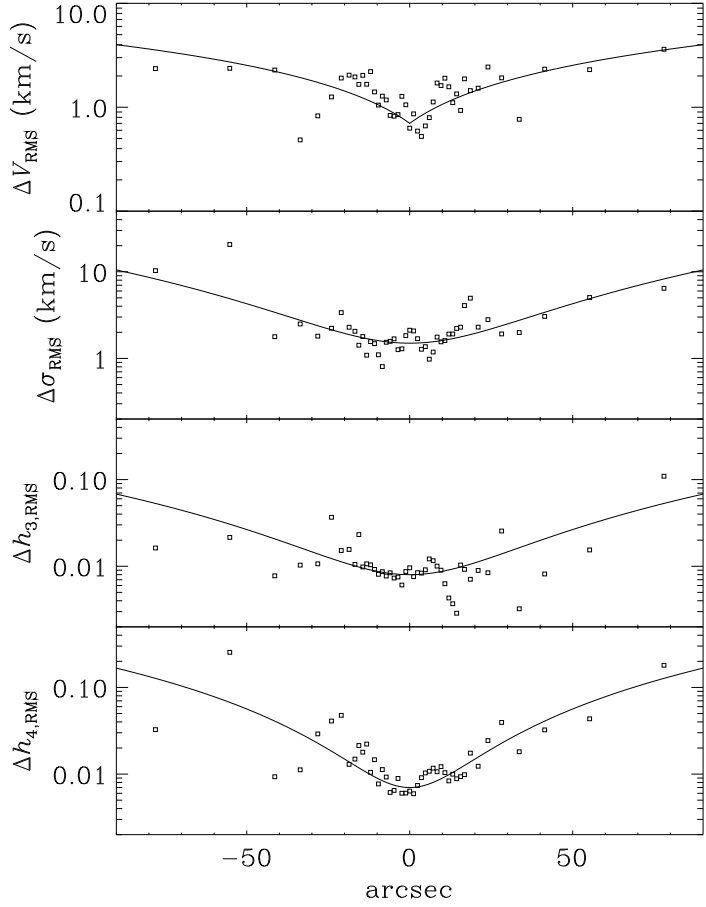}\hfill}
\caption{\footnotesize
Systematic error due to template mismatch. Differences
in the kinematic parameters obtained using two different composite templates
are plotted against radius. Plotted points are the RMS over the four
position angles. Smooth curves indicate fitting functions given in
equation (\protect{\ref{e.systemplate}}).
\label {f.systemplate}}
\end{figure}

\begin{figure}[t]{\hfill\epsfxsize=3.7in\epsfbox{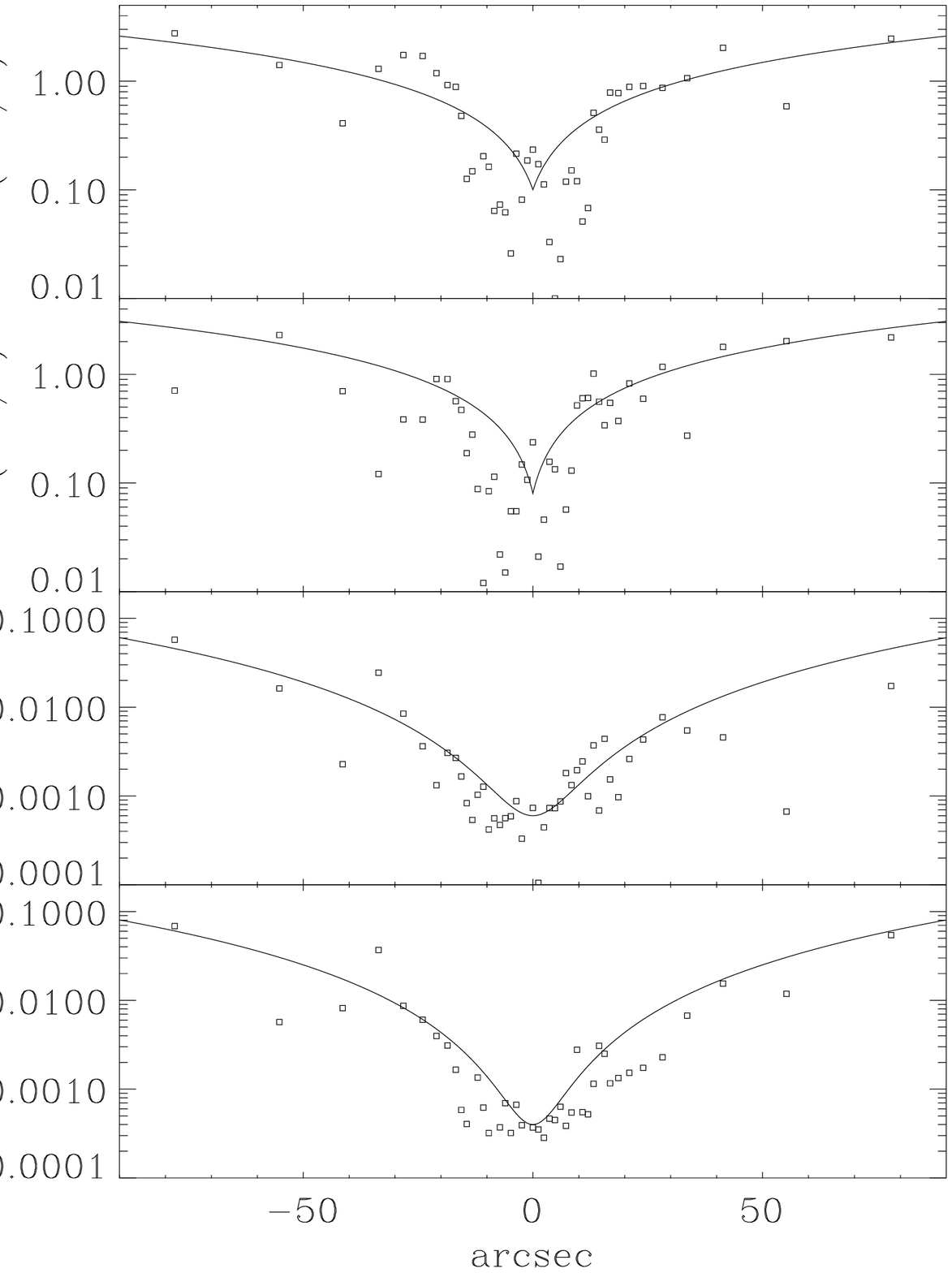}\hfill}
\caption{\footnotesize
Systematic error due to sky subtraction. Differences in
the results obtained using two different sky subtractions are plotted
against radius, for PA 70 only. Smooth curves indicate the fitting
functions given in equation (\protect{\ref{e.syssky}}).
\label {f.syssky}}
\end{figure}

\begin{figure}[t]{\hfill\epsfxsize=4.1in\epsfbox{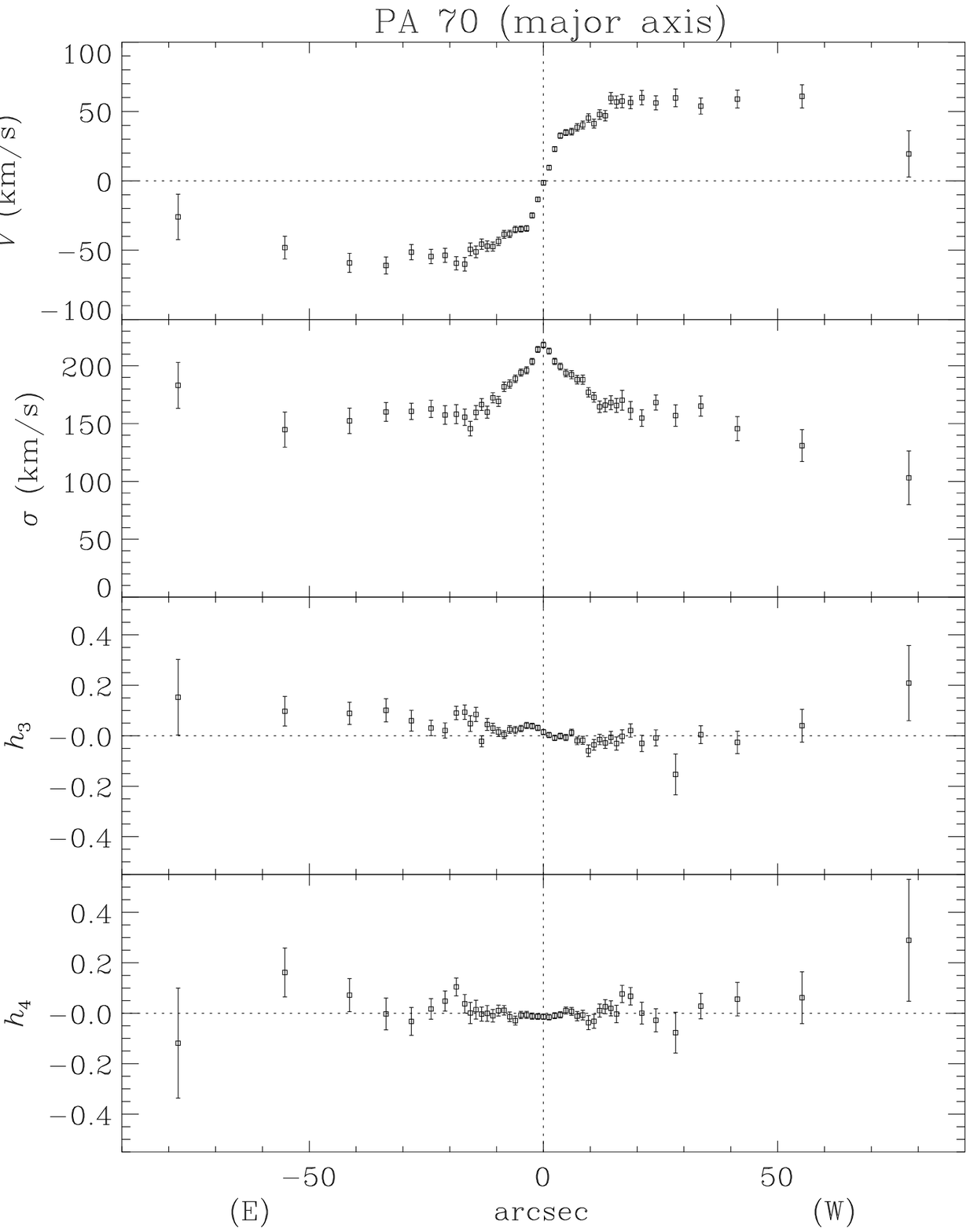}\hfill}
\caption{\footnotesize
Kinematic profiles for NGC 3379. $V$, $\sigma$, $h_3$, and $h_4$ are
the parameters in the truncated Gauss-Hermite expansion for the line-of-sight
velocity distribution, equation (\protect{\ref{e.losvd}}). (a) PA 70
(major axis).
\label {f.kinematicprofiles}}
\end{figure}

\subsection{Systematic Errors\label{s.systematic}}

Formal uncertainties on the results presented in Sec.\ \ref{s.results}
below are obtained
from the covariance matrix returned by the XC algorithm. But
we also need to estimate the dominant systematic
errors, associated with sky subtraction and
template mismatch. To accomplish this, we carried out parallel reductions of
the data using the ``naive sky'' subtraction described in Sec.
\ref{s.initial}, and using composite templates generated from a different
set of library spectra.

Figure \ref{f.systemplate} shows the differences in the
kinematic parameters obtained using the different library spectra
for the composite templates, plotted against radius. The plotted points
give the root-mean-square differences, with the mean taken over the four
position angles. We have fitted these data by eye with the following
functions:\\
\parbox{5.9in}{
\begin{eqnarray*}
\Delta V_{\rm rms} &=& 0.037 |R| + 0.70, \\
\Delta \sigma_{\rm rms} &=& 0.0011 R^2 + 1.5, \\
\Delta h_{3,\rm rms} &=& 7.4 \times 10^{-6} R^2 + 0.008, \\
\Delta h_{4,\rm rms} &=& 2.0 \times 10^{-5} R^2 + 0.007,
\end{eqnarray*}}\hfil
\parbox{.3in}{
\begin{eqnarray}
\label{e.systemplate}
\end{eqnarray}}\\
where $\Delta V_{\rm rms}$ and $\Delta \sigma_{\rm rms}$ are given in
$\kms$ and $R$ is in arcseconds. These fits are plotted as the smooth
curves in Figure \ref{f.systemplate}.

The corresponding differences between the adopted sky subtraction and
the ``naive sky'' approach are shown in Figure \ref{f.syssky}. Here the
analysis has been repeated only for PA 70, so there is no averaging over
position angle. The smooth curves show the fitting functions, given by
\parbox{5.9in}{
\begin{eqnarray*}
\Delta V &=& 0.028 |R| + 0.10, \\
\Delta \sigma &=& 0.033 |R| + 0.08, \\
\Delta h_3 &=& 7.4 \times 10^{-6} R^2 + 0.0006, \\
\Delta h_4 &=& 9.9 \times 10^{-6} R^2 + 0.0004.
\end{eqnarray*}}\hfil
\parbox{.3in}{
\begin{eqnarray}
\label{e.syssky}
\end{eqnarray}}\\
Comparison of the figures shows that template mismatch dominates 
sky subtraction in the systematic error budget by more than an order
of magnitude in the bright center of the galaxy, but by only factors
of order unity at the slit ends.

The final error bars given in Table 1 and the figures represent the
formal internal errors from the XC code added in quadrature with the
contributions from equations (\ref{e.systemplate}) and (\ref{e.syssky}).

\section{Results\label{s.results}}

\subsection{Parametric Profiles\label{s.parametric}}

Kinematic profiles along the four sampled PAs are shown in Figure
\ref{f.kinematicprofiles}a--d. For each PA, we plot the
Gauss-Hermite parameters $V$, $\sigma$, $h_3$, and $h_4$, which are also
listed along with their uncertainties in columns 2 -- 9 of Table 1. Remember
that only when $h_3=h_4=0$ are $V$ and $\sigma$ equal to
the true mean and dispersion, $\langle v \rangle$ and
$(\langle v^2 \rangle - \langle v \rangle^2)^{1/2}$; we will recover the
latter quantities in Sec.\ \ref{s.corrected}. In the plotted
rotation curves, we have subtracted a systemic velocity $V_{\rm sys}
= 911.9 \pm 0.2 \kms$, which has been determined
from pure Gaussian fits to the broadening functions, 
averaging pairs of points in the resulting $V$ profiles on opposite
sides of the center.

\addtocounter{figure}{-1}
\begin{figure}[t]{\hfill\epsfxsize=4.1in\epsfbox{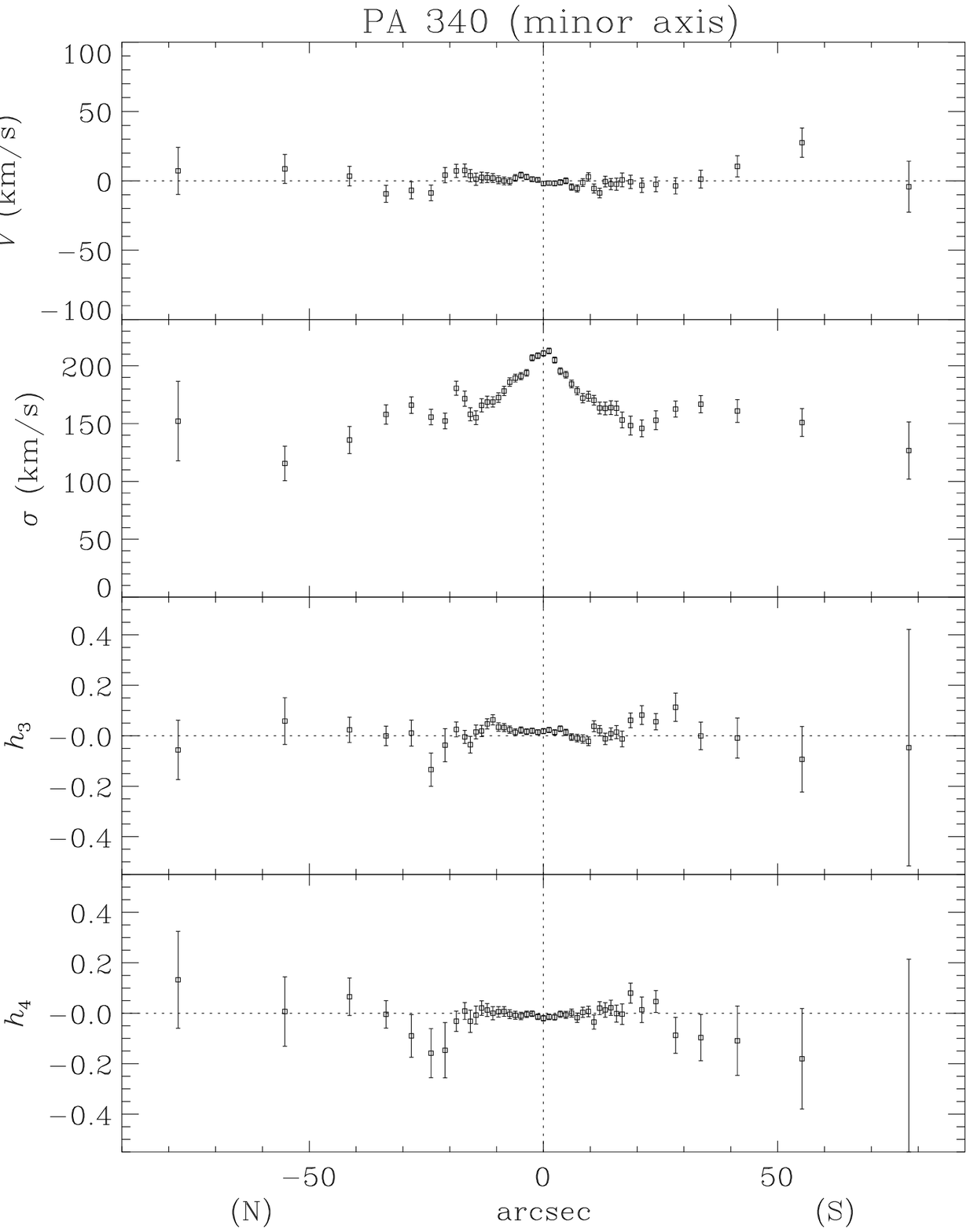}\hfill}
\caption{\footnotesize 
(b) As in (a), but for PA 340 (minor axis).}
\end{figure}

\addtocounter{figure}{-1}
\begin{figure}[t]{\hfill\epsfxsize=4.1in\epsfbox{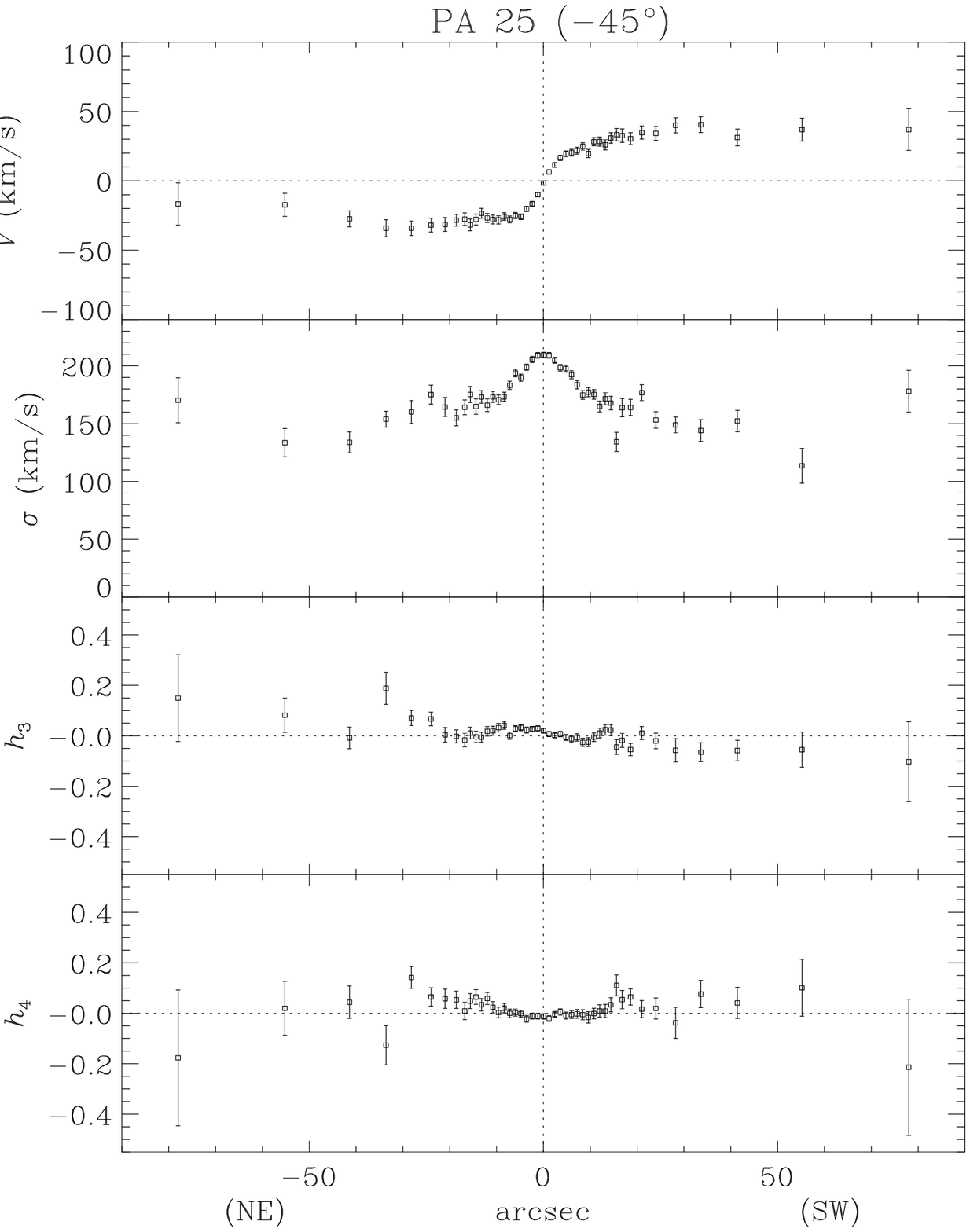}\hfill}
\caption{\footnotesize
(c) As in (a), but for  PA 25.}
\end{figure}

\addtocounter{figure}{-1}
\begin{figure}[t]{\hfill\epsfxsize=4.1in\epsfbox{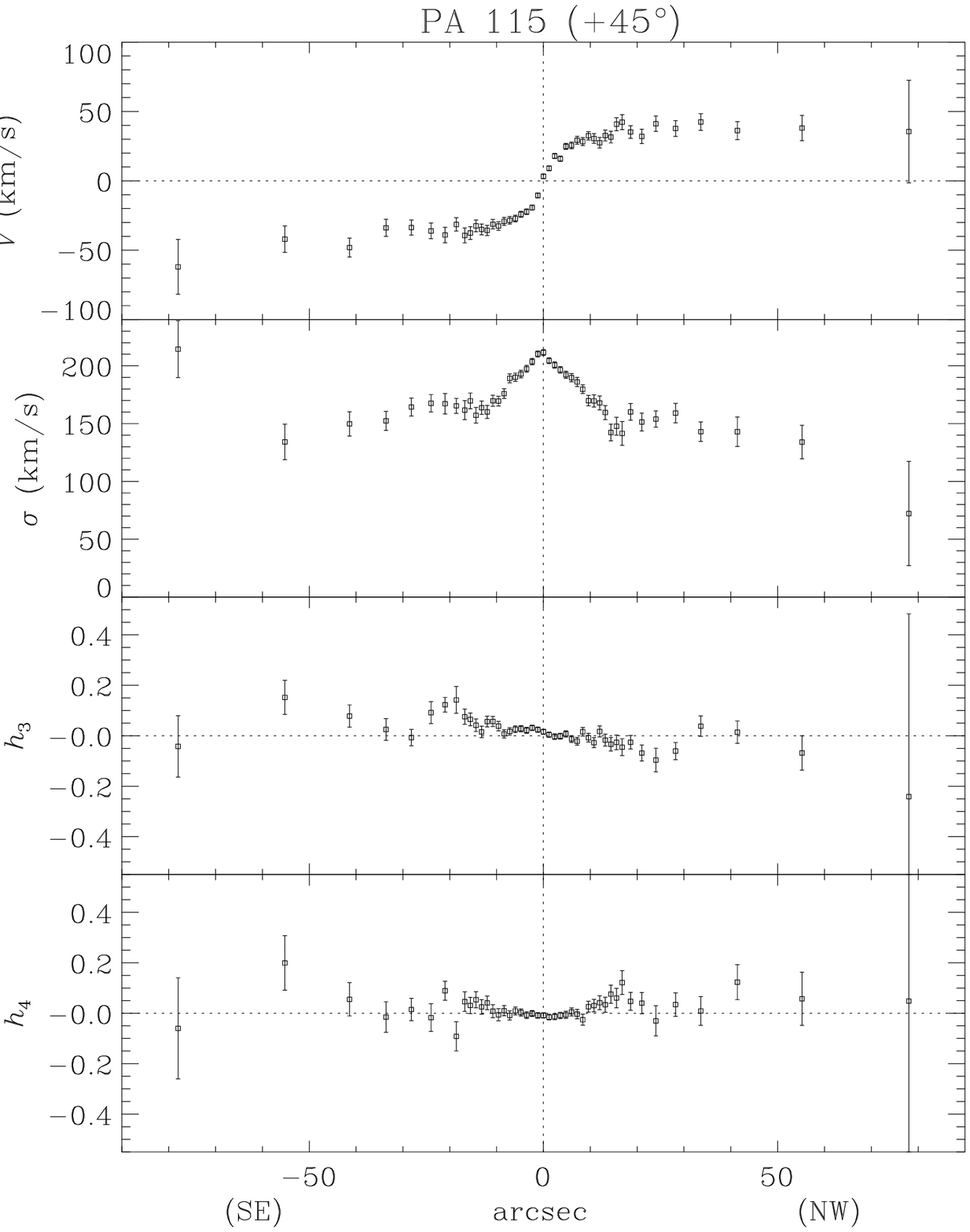}\hfill}
\caption{\footnotesize
(d) As in (a), but for PA 115.}
\end{figure}

The issue of possible minor-axis rotation is settled fairly clearly
by Figure \ref{f.kinematicprofiles}b. PA 340 shows only very weak rotation,
at about the $3\kms$ level, inside $12\arcsec$. Doubling the radial
bin size and folding about the origin to improve $S/N$, we find no detectable
rotation above $6\kms$ from $12\arcsec$ to $50\arcsec$ or above $16\kms$ out
to $90\arcsec$ (95\% confidence limits). The maximum rotation speed of
approximately $60 \kms$ is found on the major axis, and intermediate
speeds are found on the diagonal PAs.
The most striking features of the kinematic profiles, however, are
the sharp bends in the major-axis rotation curve at $4\arcsec$ and
$17\arcsec$, and the comparably sharp inflections near $15\arcsec$ in all
of the $\sigma$ profiles. These kinks are invisible in the earlier data,
which have insufficient spatial resolution and kinematic accuracy to
reveal them (cf.\ Fig. \ref{f.otherdata} below).

Careful inspection of the $h_3$ and $h_4$ profiles suggests features
coincident with the kinks in
$V$ and $\sigma$, though this is difficult to see because the Gauss-Hermite
terms have proportionally larger
error bars. To improve the statistics, we have combined the data in Figure
\ref{f.kinematicprofiles} to create composite, azimuthally averaged
radial profiles. The mean $V$ profile is scaled to the major axis amplitude
by multiplying the PA 25 and PA 115 data by factors of $1.65$ and $1.41$,
respectively, before folding (antisymmetrizing) about the center and averaging;
the minor axis (PA 340) data is omitted from
the composite $V$ and $h_3$ profiles. Since we see no significant
differences with PA in the $\sigma$ and $h_4$ profiles, for these we simply
symmetrize and average all 4 PAs with no scaling.

\begin{figure}[t]{\hfill\epsfxsize=3.7in\epsfbox{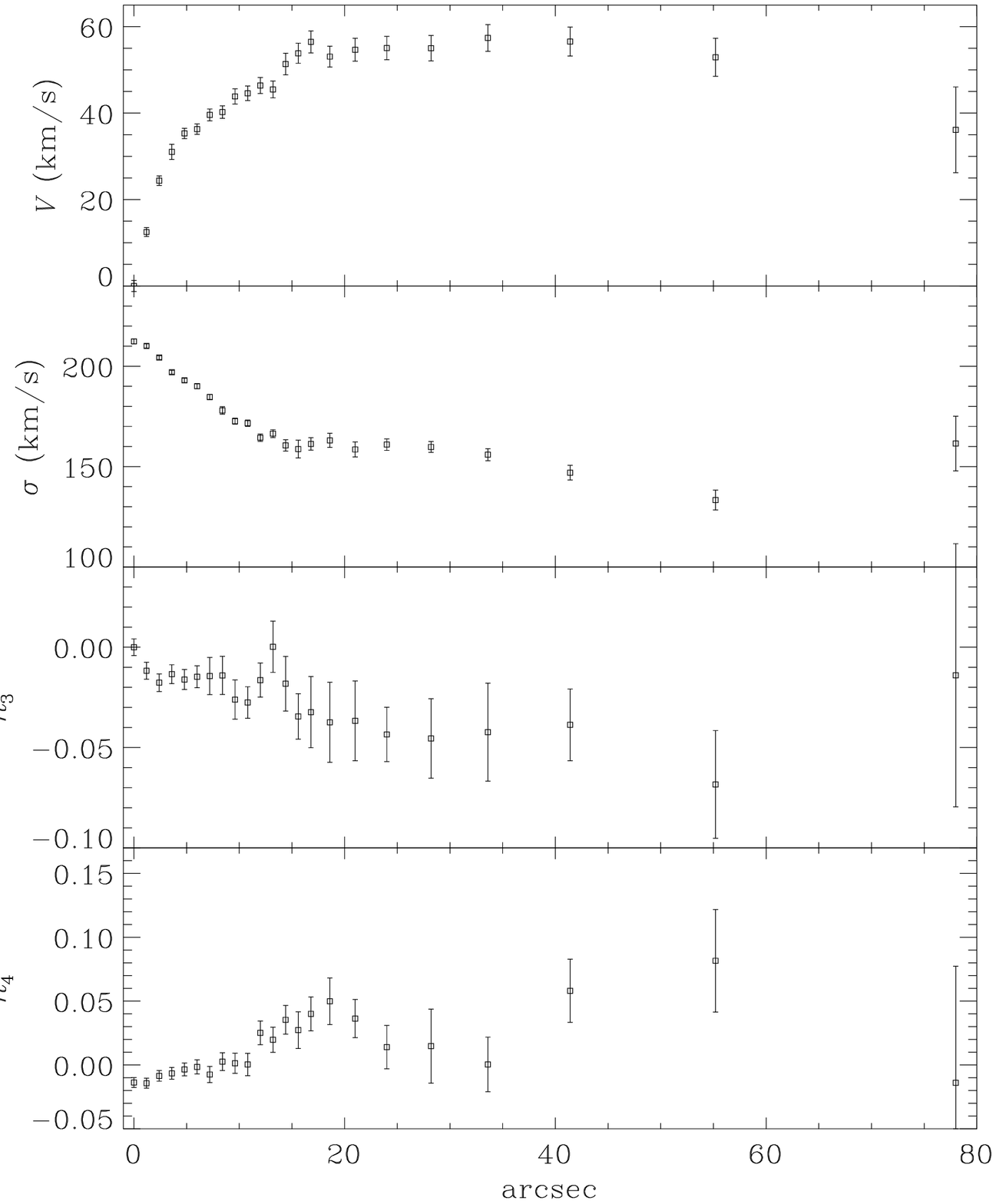}\hfill}
\caption{\footnotesize
Composite radial kinematic profiles derived from all four slit PAs, as
described in Sec.\ \protect{\ref{s.parametric}}.
\label{f.radialprofiles}}
\end{figure}

The resulting radial profiles are shown in Figure \ref{f.radialprofiles}.
The shapes of the $V$ and $\sigma$ profiles are clarified, particularly
the almost piecewise-linear form of the rotation curve and the sudden
transition in $\sigma(R)$ near $15\arcsec$. We also see a small bump
at $13\arcsec$ in the $h_3$ profile. $V$ and $h_3$ are of opposite sign
at all radii, consistent with the usual sense of skewness. The $h_4$
profile shows a clear positive gradient out to $18\arcsec$, where it
turns over, then gradually increases again beyond about $35\arcsec$.
Positive $h_4$ indicates an LOSVD that is more ``peaky'' and has longer
tails than a Gaussian.
The change of sign of $h_4$ in the inner $7\arcsec$ or so should not be
taken too literally, since a constant offset in $h_4$ is an expected artifact
of template mismatch.

The clustering of interesting kinematic features in the region from
$13\arcsec$ to $20\arcsec$ is intriguing. In order of increasing $R$, we see
a drop in the skewness of the LOSVD, an abrupt flattening in the dispersion
and rotation curves, and a local maximum in the LOSVD ``peakiness.''
This is, moreover, a photometrically interesting region. Capaccioli et
al.\ \markcite{Cap90}(1990) find residuals from the best-fitting $r^{1/4}$ law
as large as $0.2$ mag; the logarithmic slope of the $B$-band surface brightness
profile peaks at $18\arcsec$. Evidently, this range of radii marks a very
important transition in the galaxy.

\subsection{Corrected Mean Velocities and Dispersions\label{s.corrected}}

For dynamical models based on the low-order moment (continuity and Jeans)
equations, it is important to have the true mean and dispersion, rather than
the Gauss-Hermite parameters $V$ and $\sigma$. We can calculate these
quantities and their associated errors using equations (5)--(7)
of SSC, which are based on the treatment of van der Marel \& Franx
\markcite{vdMF93}(1993).

First, however, we must determine whether including
the $h_3$ and $h_4$ terms actually results in a statistically significant
improvement to the estimate of the moments $\langle v \rangle$ and
$\langle v^2 \rangle$ from the data. We found above that $h_3$ and $h_4$
are generally small, and since their error bars grow with radius, it
is not obvious {\em a priori\/} that correcting $\langle v \rangle$ and
$\langle v^2 \rangle$ for these terms---and increasing the error
bars accordingly---will necessarily give more robust estimates than
simply assuming a Gaussian LOSVD. Therefore, we examine the distribution
of chi-square values, $\chi ^2_3$ and $\chi^2_5$, obtained from, respectively,
three-parameter (Gaussian) and five-parameter (Gauss-Hermite) fits to the
broadening functions. We find that the differences $\Delta\chi = \chi^2_3 -
\chi^2_5$ are significant only for $R < 4\arcsec$. For the rest of the
data, the distribution $F(\chi^2_3)$ is completely consistent with a
chi-square distribution with the appropriate number of degrees of
freedom, if our original estimates for the noise in the galaxy spectra
are scaled up by a factor of $1.17$. The noise in each
spectrum is estimated by differencing the spectrum with a smoothed
version of itself; and it is certainly believable that this procedure
could underestimate the actual noise level by 17\%. All of
the results in this paper have been computed including this correction
to the noise.

\begin{figure}[t]{\hfill\epsfxsize=3.7in\epsfbox{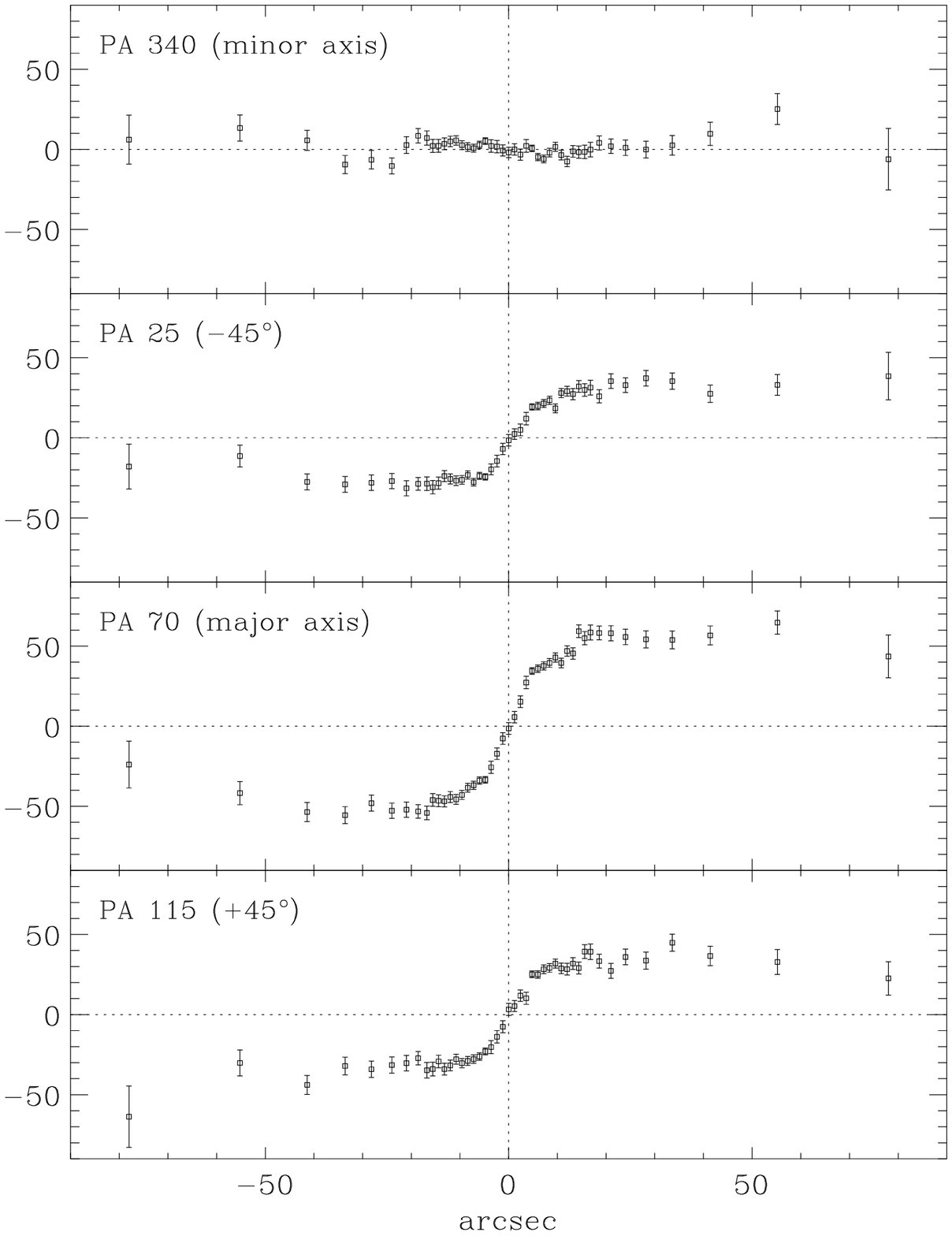}\hfill}
\caption{\footnotesize
(a) Mean velocity profiles, corrected for the
non-Gaussian terms in the LOSVD as described in
Sec.\ \protect{\ref{s.corrected}}.
\label{f.correctedprofiles}}
\end{figure}

The adopted mean velocity and dispersion profiles are shown in Figure
\ref{f.correctedprofiles}a--b and listed in the last 4 columns of Table
1. For $R<4\arcsec$, we use the
results of the Gauss-Hermite fits, corrected for $h_3$ and $h_4$. To
avoid propagating the residual effects of template mismatch,
we have applied a constant offset to the $h_3$ profile on each PA so as to
shift the central value to zero.
For larger radii we adopt the $V$ and $\sigma$ values from pure-Gaussian
fits. We are not saying that the LOSVD {\em is\/} Gaussian
beyond $4\arcsec$, merely that the most reliable estimates of the mean and
dispersion come from the Gaussian fit. Since the corrections are all
small, the corrected rotation curves resemble the $V$ profiles in Figure
\ref{f.kinematicprofiles}, including the very weak minor-axis rotation, 
the sharp kinks in the major-axis profile, and a slightly higher
rotation speed on PA 115 diagonal than on PA 25. The $h_4$ corrections
to the dispersion flatten out the central gradient slightly and have
little effect on the rest of the profiles.

\addtocounter{figure}{-1}
\begin{figure}[t]{\hfill\epsfxsize=3.7in\epsfbox{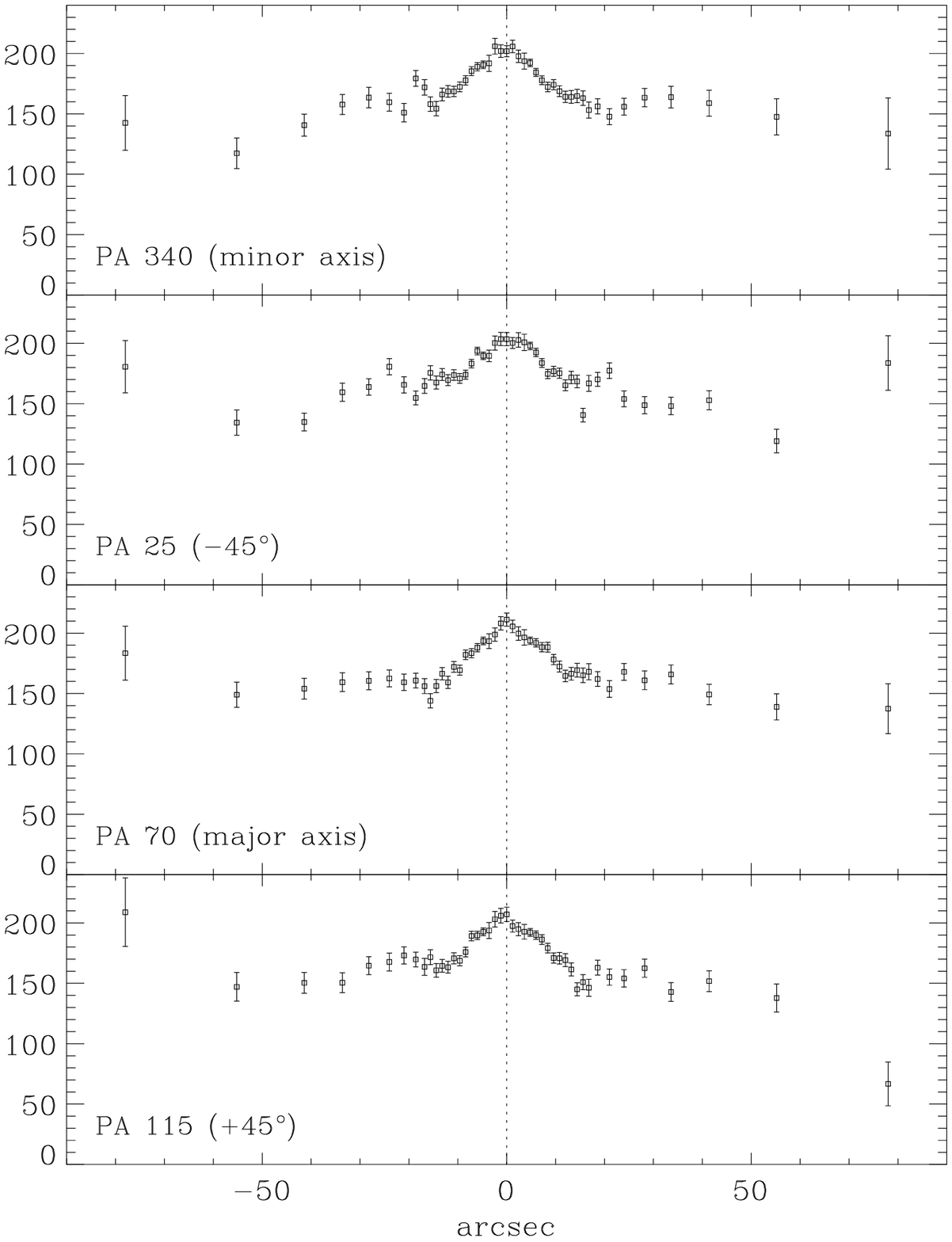}\hfill}
\caption{\footnotesize
(b) velocity dispersion profiles, as in (a).}
\end{figure}

\subsection{Reconstructed Two-Dimensional Fields\label{s.twod}}

With multiple-PA sampling, we can create Fourier reconstructions
of the two-dimensional kinematic fields from the profiles in Figure
\ref{f.correctedprofiles}. Our $45\arcdeg$ spacing lends
itself to a representation of the form
\beq
f(R,\theta) = C_0 + \sum_{i=1}^4 \left(C_i \cos m \theta
	 + S_i \sin m \theta\right),
\eeq
where the coefficients are all functions of $R$, and $S_4\equiv 0$ if we let
one of our sampled PAs correspond to
$\theta=0$. An explicit expression for the reconstructed velocity field in
terms of the measured velocities is given in equation (9) of SSC; the
corresponding expression for the dispersion field has the same form since no
particular symmetry is assumed. To reduce the noise in the 2-D fields, we
interpolate and smooth the 1-D profiles using a smoothing spline
(Green \& Silverman \markcite{GrS94}1994) before computing the reconstructions.

\begin{figure}[t]{\hfill\epsfxsize=3.0in\epsfbox{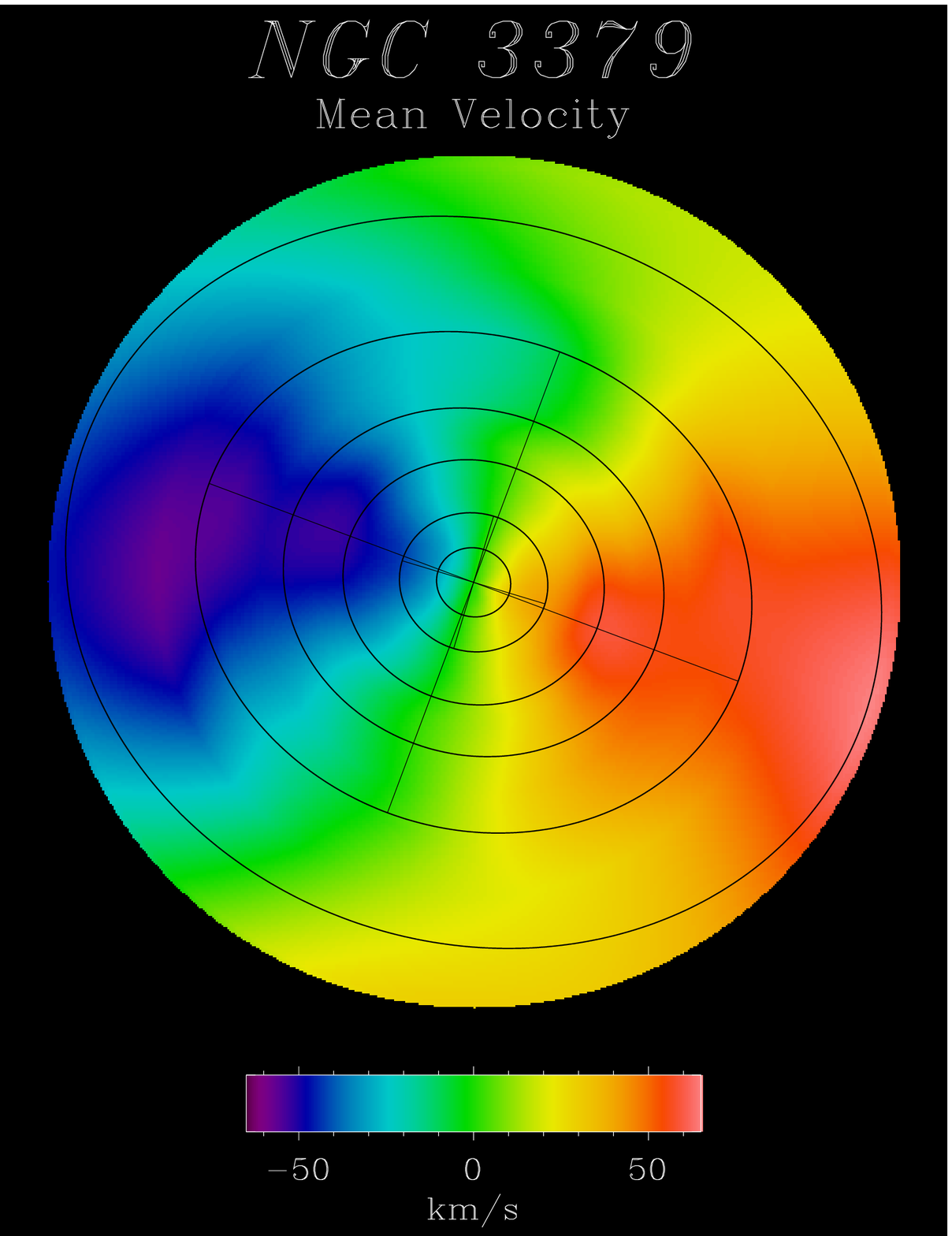}\hfill}
\caption{\footnotesize
(a) Fourier reconstruction of the mean velocity field.
Ellipses show isophotes from Peletier et al.\ (1990); major and minor axes
drawn for two isophotes indicate the magnitude of the isophotal twist. The
plotted region is $56\arcsec$ in radius.
Notice the twist of the kinematic major axis (line joining the extreme
velocities at each radius) in the direction opposite to the isophotal
twist.
\label{f.vfield}}
\end{figure}

The resulting velocity and dispersion fields are shown in Figure
\ref{f.vfield}a--b. The plotted region is $56\arcsec$ in radius, which
omits only the outermost points on each PA. Black ellipses show
representative isophotes, as fitted by Peletier et al.\ \markcite{Pel90}(1990);
we have drawn the principal axes for two of the isophotes to
indicate the modest photometric twist in the galaxy. In Figure
\ref{f.vfield}a, note the rotation of the kinematic major axis away from the
photometric major axis for $R \gtrsim 30\arcsec$. This rotation is due
in roughly equal measure to the $4\arcdeg$ isophotal twist and to a $\sim
5\arcdeg$ kinematic twist of the velocity field in the opposite direction.
Figure \ref{f.vfield}b
nicely illustrates the steep central rise in the dispersion, as well
as the quite flat profile outside of $15\arcsec.$ The odd structure with
apparent 3-fold symmetry is almost certainly not real; however, the
azimuthally averaged profile does show a very weak ``hump'' near
$20\arcsec$, which would be consistent with a ring of slightly higher
dispersion at around this radius.

\addtocounter{figure}{-1}
\begin{figure}[t]{\hfill\epsfxsize=3.0in\epsfbox{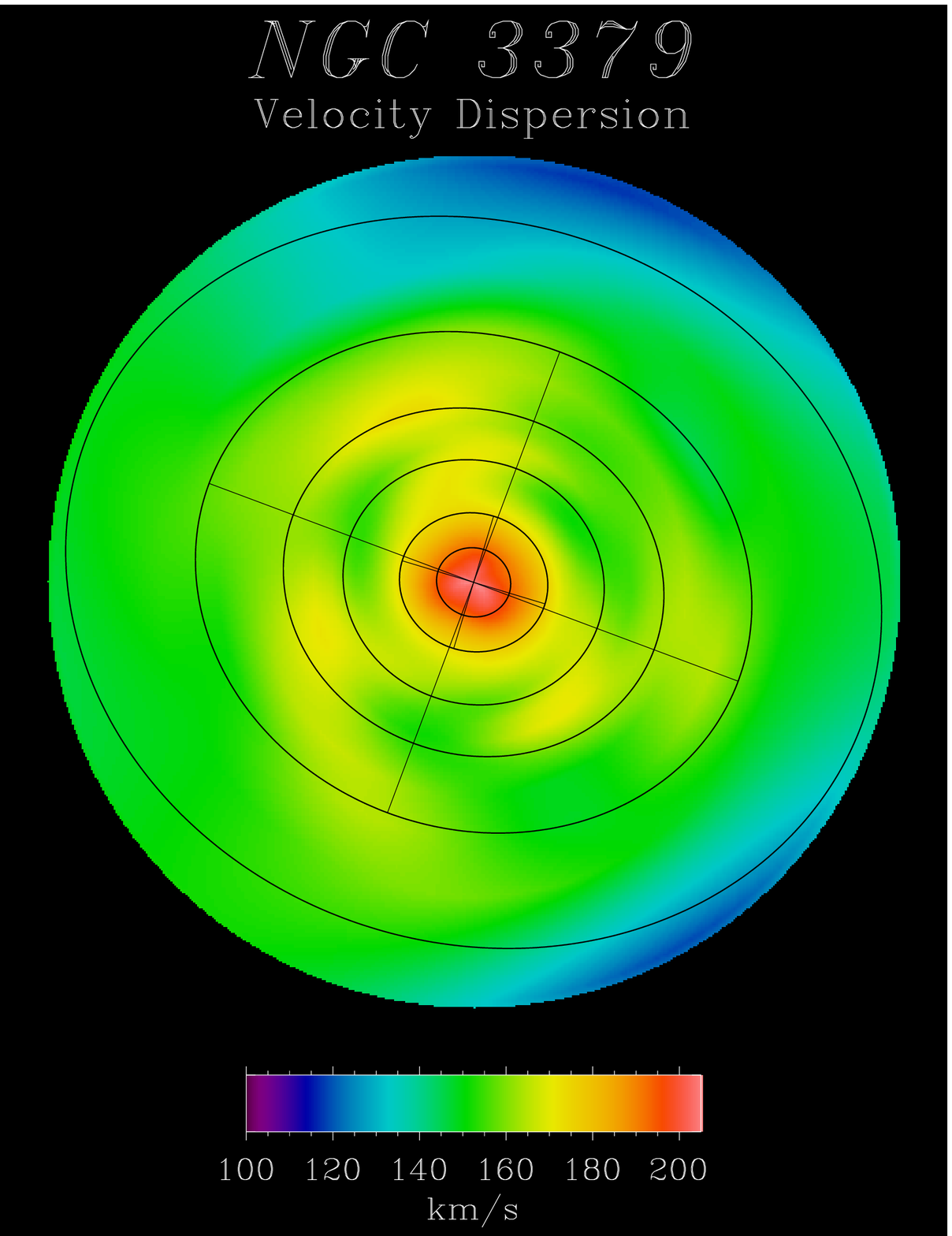}\hfill}
\caption{\footnotesize
(b) Fourier reconstruction of the velocity dispersion field, as in (a).}
\end{figure}

\subsection{Non-Parametric LOSVDs\label{s.nonparametric}}

\begin{figure}[t]{\hfill\epsfxsize=3.0in\epsfbox{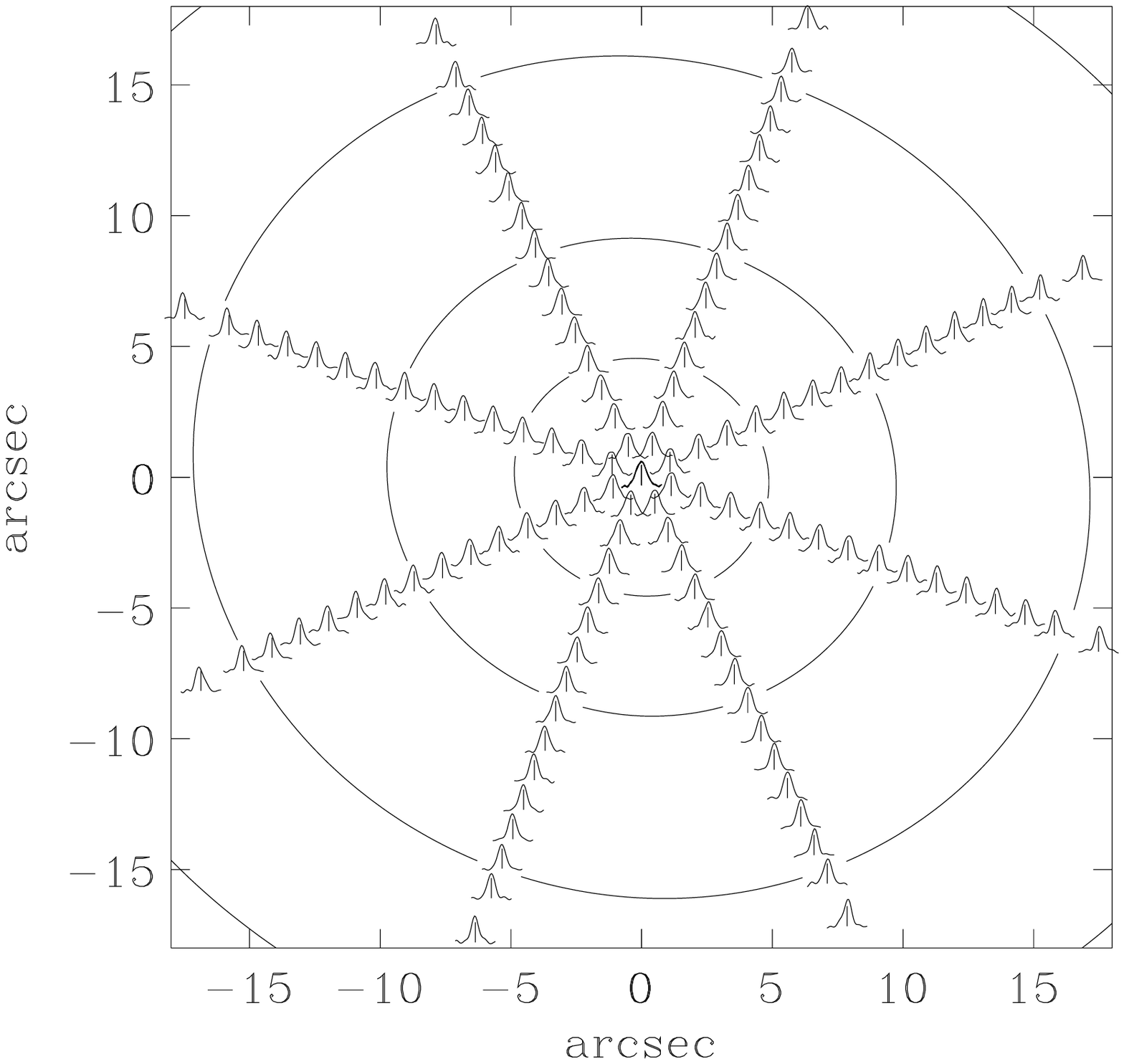}\hfil\epsfxsize=3.0in\epsfbox{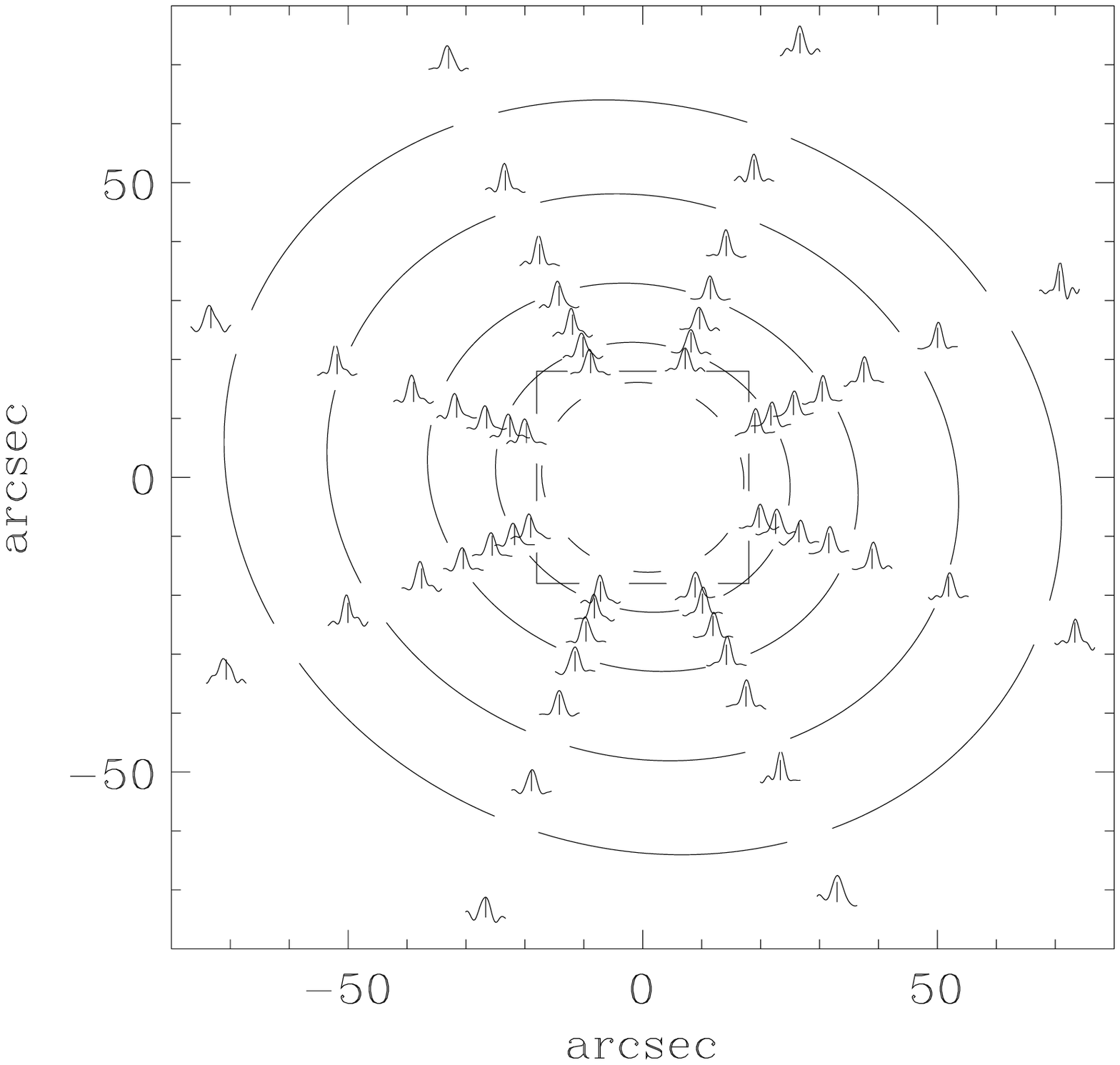}\hfill}
\caption{\footnotesize
LOSVDs obtained by Fourier Correlation Quotient method, plotted at the sampled
positions on the sky. Vertical lines indicate the systemic velocity.
Representative isophotes are shown for orientation. (a) Inner region; (b)
outer region. Square in (b) shows the area plotted in (a).
\label{f.fcqplot}}
\end{figure}

Non-Parametric LOSVDs derived by the FCQ method are plotted in Figure
\ref{f.fcqplot}a--b, at the appropriate positions on the sky. Representative
isophotes are shown for orientation; in each little profile, the vertical line
marks the systemic velocity. Consistent with the results of Sec.
\ref{s.parametric}, one can see that nowhere is the LOSVD strongly
non-Gaussian. Careful inspection, however, does show a very modest
skewness in the usual sense along the inner major axis, and a
tendency for the LOSVD to be slightly sharper-peaked at large radii.

\pagebreak

\section{Discussion\label{s.discussion}}

\nopagebreak

\subsection{Comparison with Previous Work}

\begin{figure}[t]{\hfill\epsfxsize=5.0in\epsfbox{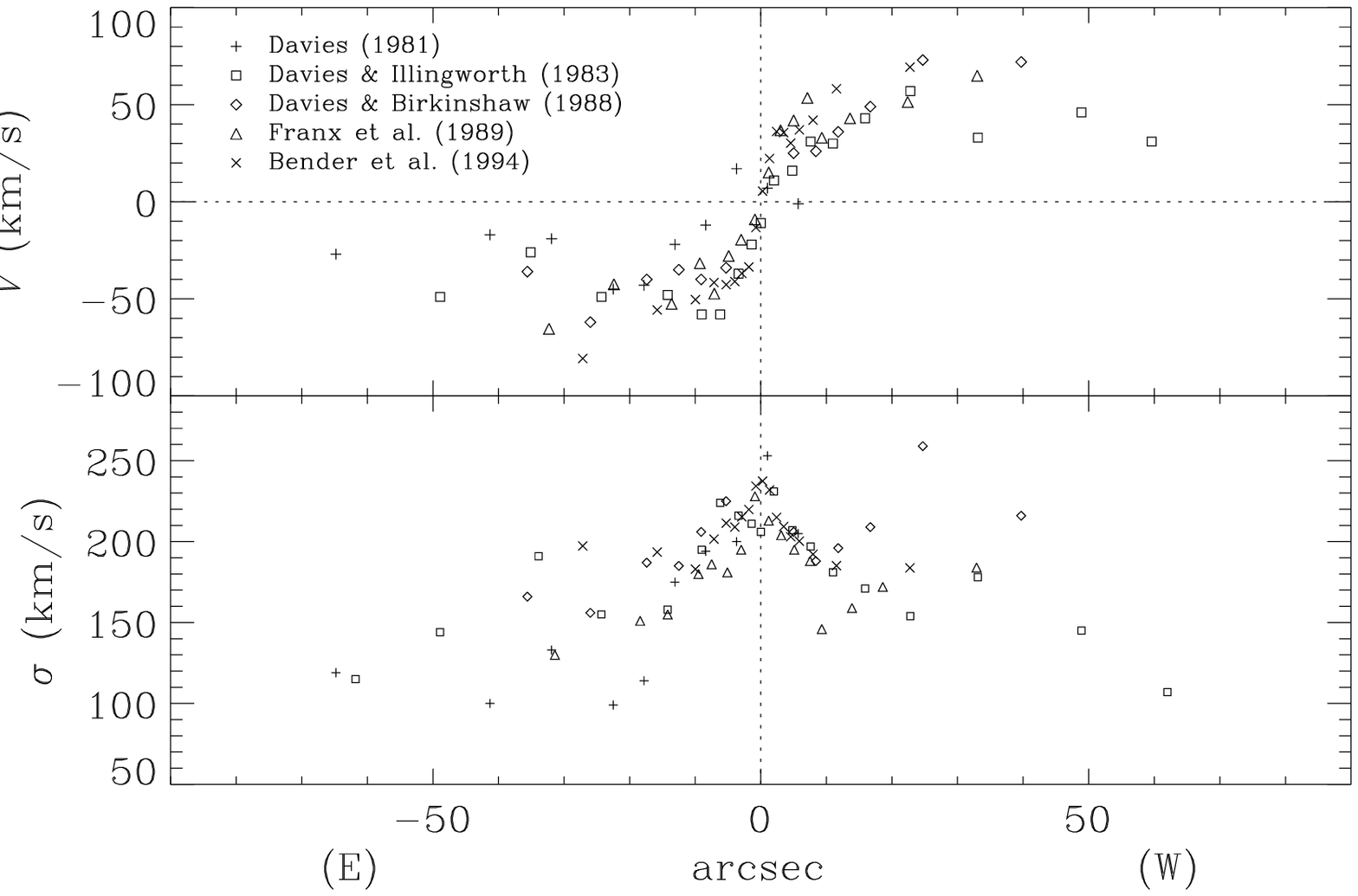}\hfill}
\caption{\footnotesize
Major-axis (or near-major-axis) $V$ and $\sigma$ profiles from previous
authors, for comparison with the top two panels of Fig.\ 
\protect{\ref{f.kinematicprofiles}}a. Error bars have been omitted for clarity.
\label{f.otherdata}}
\end{figure}

Kinematic data for NGC 3379 have been published previously by
Sargent et al. \markcite{Sar78}(1978), Davies \markcite{Dav81}(1981),
Davies \& Illingworth \markcite{DaI83}(1983), Davies \& Birkinshaw
\markcite{DaB88}(1988), Franx et al.\ \markcite{FIH89}(1989), and
Bender, Saglia, \& Gerhard
\markcite{BSG94}(1994). Major axis $V$ and $\sigma$ profiles from all but
the first of these studies are plotted in Figure \ref{f.otherdata}.
Comparison with the top two panels of Fig.\ \ref{f.kinematicprofiles}a
shows that the present data are largely consistent with the earlier results,
but reveal structure that could not be seen in the earlier data. With the
benefit of hindsight, one can discern a change of slope in
$\sigma(R)$ near $20\arcsec$, but this feature is quite murky except in
the Bender et al.\ data. A steep
decline in dispersion is noted by Davies \markcite{Dav81}(1981), though
his mean dispersion of $114\kms$ outside of $15\arcsec$ is not reproduced
in the later work. Davies and Illingworth \markcite{DaI83}(1983)
conclude that the overall gradient is consistent with constant $M/L$.
Their data also show a shallow hump in $\sigma(R)$ beyond $20\arcsec$,
but at no more than $1\sigma$ significance. Bender et
al.\ \markcite{BSG94}(1994) obtain a $\sigma$ profile with a local minimum
of approximately $185\kms$ near $10\arcsec$, rising again to about
$200\kms$ at $27\arcsec$, their outermost data point.
This is somewhat inconsistent with our results, though not alarmingly so.
Their dispersions seem to be systematically $\sim 20\kms$ higher than ours,
which could easily be caused by their use of a single template star not matched
to the galactic spectrum.
They detect a sharp bend in the rotation curve at $4\arcsec$, but do not
have fine enough sampling to see a second bend farther out.
Bender et al.\ also derive $h_3$ and $h_4$, obtaining a
generally featureless $h_3$ profile with $\langle h_3 \rangle \approx -0.02$
on the positive-velocity side, and a weak positive gradient in $h_4$.
This again is consistent with our results, though at significantly
coarser resolution.

The minor axis velocity data from Davies \& Birkinshaw \markcite{DaB88}(1988)
and Franx et al.\ \markcite{FIH89}(1989) are compiled in Figure 4d of
Statler \markcite{Sta94}(1994). Those data show a scatter of $18\kms$ about
a mean of $3\kms$, with a possible increase in rotation beyond $30\arcsec$.
As discussed
in Sec. \ref{s.parametric}, we obtain 95\% confidence upper limits
of $6\kms$ for $12\arcsec < R < 50\arcsec$ and $16\kms$ for $50\arcsec < R <
90\arcsec$, with a marginal detection of $\sim 3\kms$ rotation on PA 340
interior to $12\arcsec$.

\subsection{Connection with Planetary Nebulae}

\begin{figure}[t]{\hfill\epsfxsize=3.0in\epsfbox{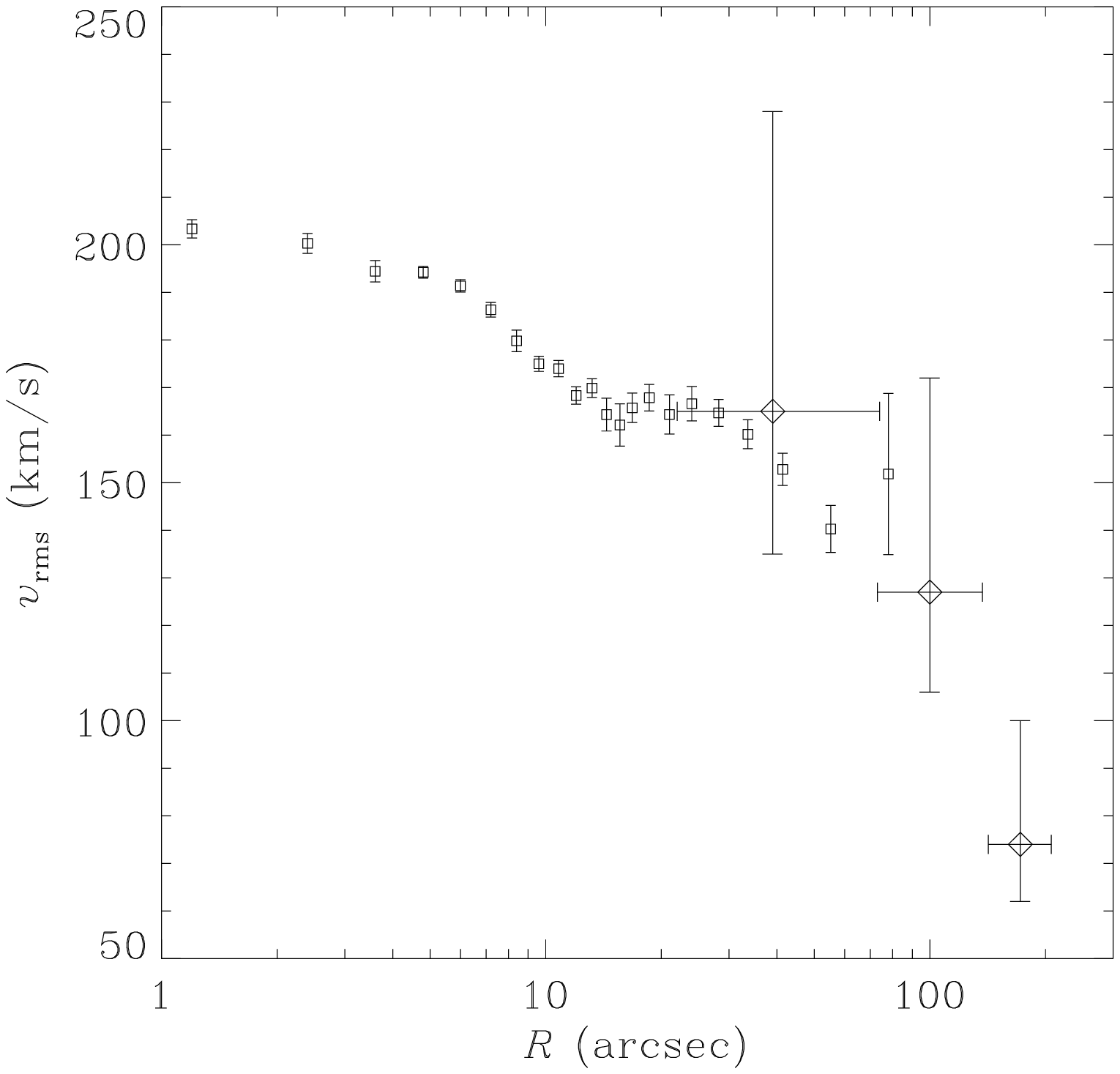}\hfill}
\caption{\footnotesize
Kinematics at small and large radii. The composite line-of-sight RMS velocity
profile of the integrated stellar light ({\em squares\/}) and the velocity
dispersion profile of planetary nebulae measured by Ciardullo et al.\ (1993)
({\em diamonds\/}) join up smoothly, to within the errors.
\label{f.pnebulae}}
\end{figure}

Ciardullo, Jacoby, \& Dejonghe \markcite{CJD93}(1993) have measured the
radial velocities of 29 planetary nebulae in NGC 3379, at radii between
$21\arcsec$ and $209\arcsec$. They find no evident rotation in the PN
population, but a clear signature of a negative radial gradient in the
RMS velocity. Breaking their sample into 3 radial bins, they obtain the
dispersion profile plotted as the large diamonds in Figure
\ref{f.pnebulae}. To compare with these data, we compute the RMS
velocity profile for the integrated starlight by computing composite
radial profiles for the Hermite-corrected mean velocity and dispersion
profiles, and adding them in quadrature. We make no correction to the
mean velocity for any assumed inclination. The results are plotted as the
squares in Figure \ref{f.pnebulae}. To within the errors, the profiles
join smoothly. (The upward jump in our last data point is not
statistically significant.) While it remains somewhat puzzling that the
PNe should show no rotation, at least from the dispersion profile it
would seem that they are representative of the general stellar
population at smaller radii.

\nopagebreak

\subsection{Implications for Dynamics and Structure}

The double-humped RMS velocity profile plotted in Fig. \ref{f.pnebulae}
provokes a strong impression of a two-component system. At the very
least, the sharp bends in the rotation curve (Fig.\ \ref{f.radialprofiles},
top) indicate that there are special radii within the galaxy where
abrupt---though perhaps subtle---changes in dynamical structure occur. This
sort of behavior is not generally thought of as being characteristic of
elliptical galaxies. But it has, in fact, been seen before in S0's,
most notably in NGC 3115. \markcite{Cap91}CVHL have argued that these
two systems share enough photometric characteristics that they could be near
twins, seen in different orientations. Here we show that they share a
kinematic kinship as well.

\begin{figure}[t]{\hfill\epsfxsize=5.0in\epsfbox{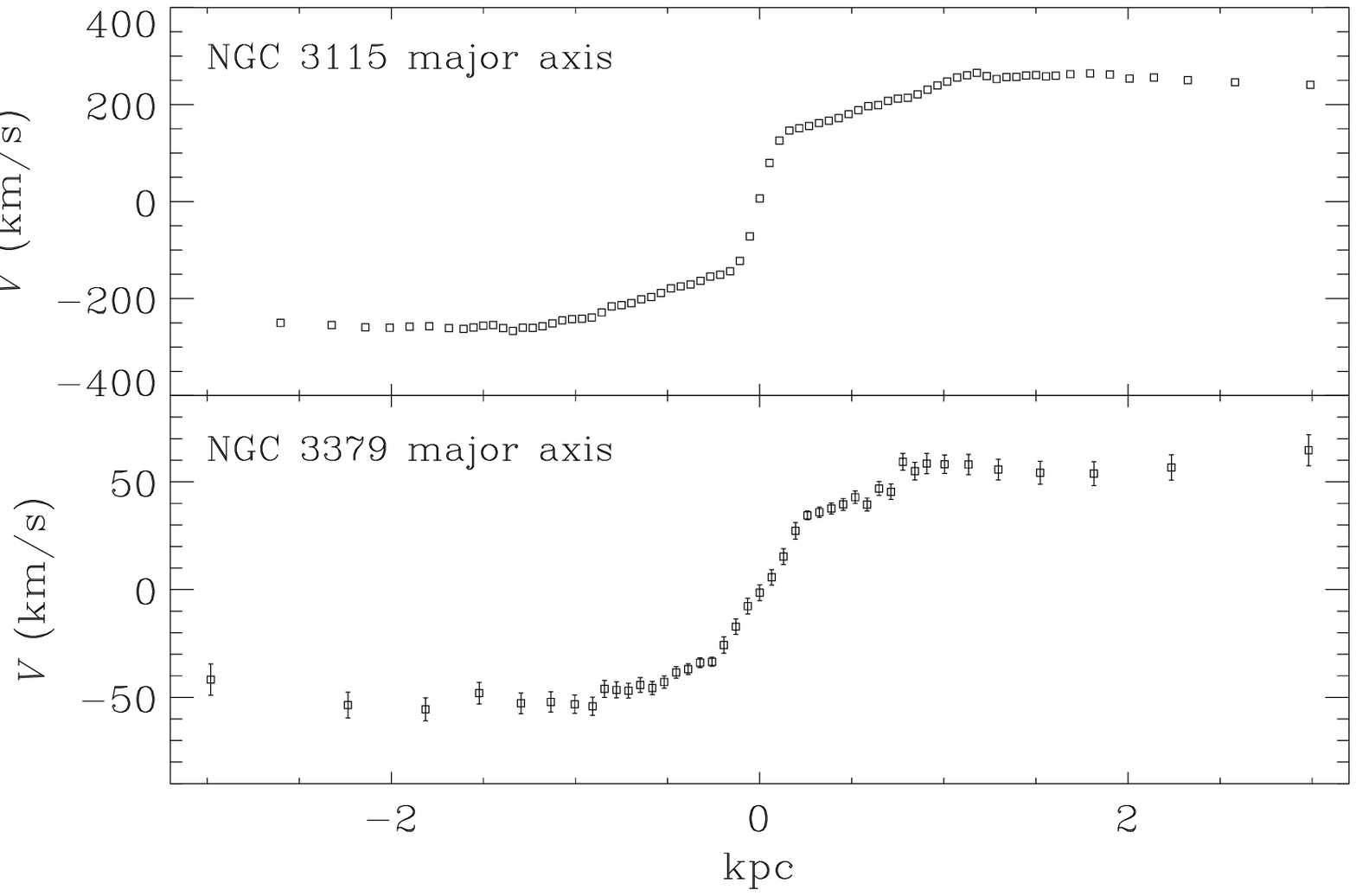}\hfill}
\caption{\footnotesize
Major axis rotation curves of NGC 3379 and NGC 3115 (Fisher 1997)
plotted on the same linear scale. Both galaxies have nearly
piecewise-linear rotation curves, with sharp bends near $0.2\kpc$ and
$1\kpc$.
\label{f.compare}}
\end{figure}

In Figure \ref{f.compare} we plot the major axis rotation curve of NGC
3115 from Fisher \markcite{Fis97}(1997), along with that of NGC 3379
on the same linear scale.
Both show a sharp inner kink $\sim 200\pc$ from the center
and an outer kink in the rough vicinity of $1\kpc$, outside of which the
rotation curve is basically flat. The similarity is striking, even
though the locations of the bends do not match exactly. NGC 3115's outer kink
coincides almost exactly with a photometric bump in the major axis
$B$-band brightness profile (Capaccioli et al.\ \markcite{CHN87}1987)
evidently related to structure in the disk. The dispersion also appears
to level off at about this radius (Bender et al.\ \markcite{BSG94}1994),
although the transition does not seem especially sharp.

\begin{figure}[t]{\hfill\epsfxsize=3.0in\epsfbox{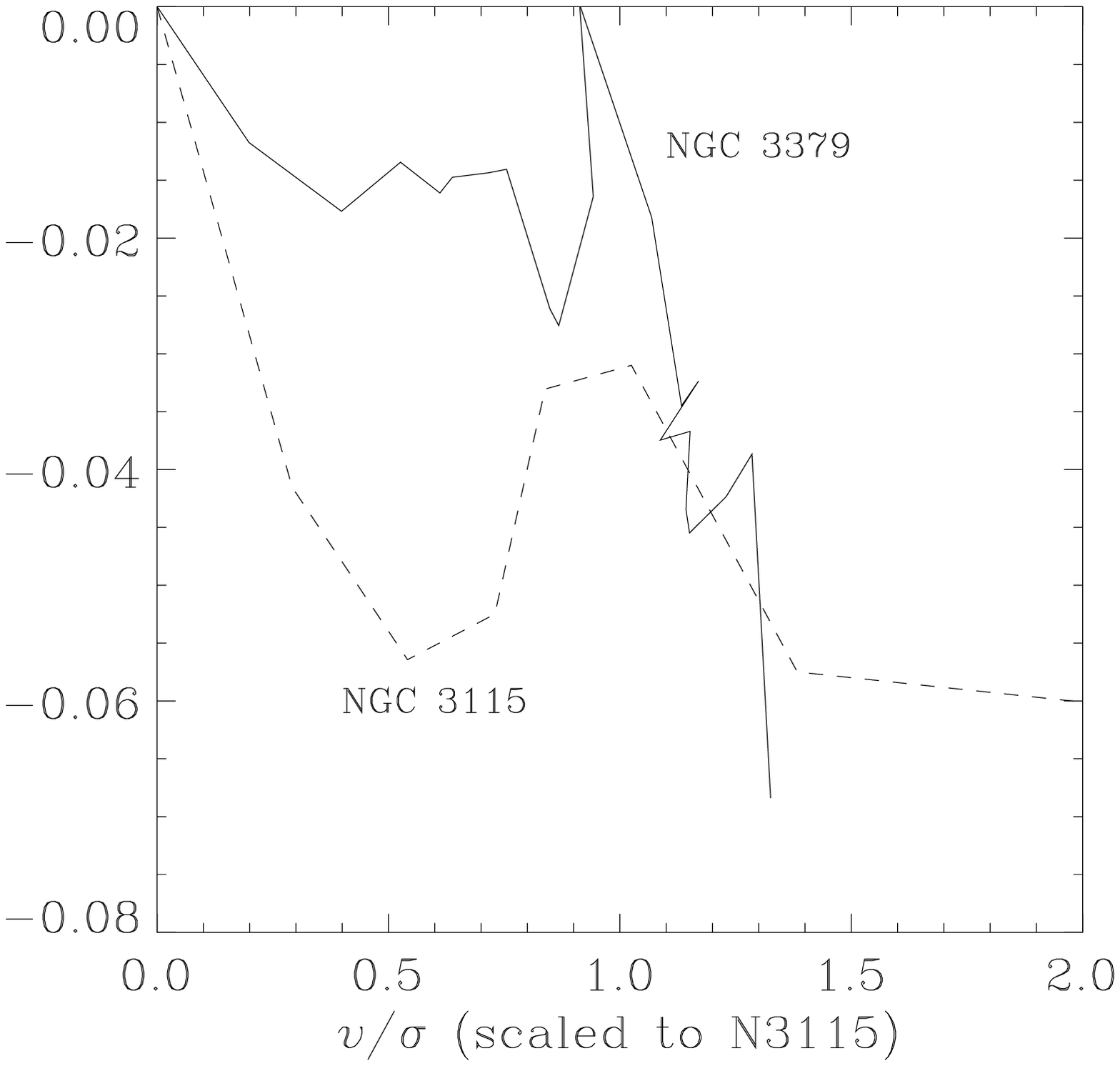}\hfill}
\caption{\footnotesize
Correlation between $h_3$ and local $v/\sigma$ at the same projected
radius, for NGC 3379 and NGC 3115. Curves have been folded
(antisymmetrized) about the center, and
the curve for NGC 3379 has been scaled to the central dispersion and
maximum rotation speed of NGC 3115. The small peaks in the $h_3$ profiles
occur at the same value of scaled $v/\sigma$.
\label{f.fisherplot}}
\end{figure}

In addition, there are similarities in the $h_3$ profiles. Fisher
\markcite{Fis97}(1997) finds a bump in $h_3$ at around $R=5\arcsec$
in NGC 3115, similar to the feature we see at $R=13\arcsec$ in NGC 3379.
Since they appear at rather different places relative to the kinks in the
rotation curves, these small bumps may be unrelated; however,
the correlation between $h_3$ and {\em local\/} $v/\sigma$ hints that
there may be a more subtle connection.
Fisher plots $h_3$ against $v/\sigma$ at the same projected radius for
a sample of 20 S0's, and finds that 10 show a distinctive N-shaped
profile through the center.
In 9 of those, $h_3$ changes sign in the legs of the N, reversing the
usual sense of skewness.  This is quite different from ellipticals
(Bender et al.\ \markcite{BSG94}1994), which tend to show only a monotonic
anticorrelation (i.e., only the middle segment of the N). In NGC 3115,
$h_3$ does not change sign more than once, but the profile turns around again 
past either end of the N-shaped segment. We plot the $h_3$ {\em vs.\/}
$v/\sigma$ profile for NGC 3115 as the dashed line in Figure
\ref{f.fisherplot}. We have taken the antisymmetric part to reduce the
noise, and plotted only the positive-velocity half, so that one sees
only the right half of the N, and the outer part ($v/\sigma >0.8$) where
the profile turns over.

To test whether NGC 3379 might plausibly be a scaled and reoriented
copy of NGC 3115, we have derived the corresponding curve for NGC 3379
from the composite radial profiles plotted in Fig. \ref{f.radialprofiles}.
We scale $\sigma$ up by a factor of $1.3$ so that the central dispersion
matches that of NGC 3115, and scale $v$ up by a factor of $4.3$ to match
the maximum speed in the flat part of NGC 3115's rotation curve. The
result is plotted as the solid line in Figure \ref{f.fisherplot}. In
terms of the scaled $v/\sigma$, the $h_3$ bump occurs in the same
place in the two galaxies.

Does this rather arbitrary scaling of $v$ and $\sigma$ correspond to
a sensible geometry? If, for simplicity, we assume an isotropic dispersion
tensor, so that the line-of-sight $\sigma$ is (at least to lowest order)
unaffected by orientation, the above scaling would require a trigonometric
factor $1.3/4.3 = \sin 18\arcdeg$ to dilute the rotation speed to the observed
value. At an inclination of $18\arcdeg$, an intrinsically E6 or E7 oblate
galaxy would be seen to have an ellipticity of $0.04$. This is a bit
rounder than the actual ellipticity of NGC 3379, which increases outward
from about $0.08$ to $0.13$ over the range of radii spanned by our data
(Peletier et al. \markcite{Pel90}1990). But the difference in apparent
shape could, in principle, be made up by a small triaxiality, so this low
an inclination is not entirely out of the question.

We would not go so far as to argue that the similarity in the $h_3$ {\em
vs.\/} $v/\sigma$ curves marks NGC 3379 as a twin of NGC 3115, or, for
that matter, as an S0 at all. There is no particular theoretical reason
to expect a bump in $h_3$ at $v/\sigma\approx 1$, no dynamical model that
predicts such a feature, and no indication that it is even present in most
S0's. But we can turn the argument around, and say that {\em if\/} it is
determined by other means that NGC 3379 is a low-inclination S0, then we
will have reason to ask what common aspect of the dynamical structure of
these two galaxies creates similar features in $h_3$ in corresponding
locations.

Heuristic arguments such as these, however, are no substitute for
dynamical modeling, which is the only proper way to determine
the true shape and dynamical structure of NGC 3379. While we leave a full
treatment of this issue to a future paper, some general discussion is
worthwhile. To lowest order, the mean velocity field of NGC 3379 is
characteristic of an oblate axisymmetric system: the highest measured
velocities are on the major axis, the minor axis rotation is near
zero, and the profiles on the diagonal slit positions are nearly the
same. Similarly, the closely aligned, almost exactly elliptical
isophotes are just what one would expect of an axisymmetric galaxy.
However, there are significant deviations, at the $\sim 10\%$ level,
from the pure axisymmetric signature, which appear as an isophotal twist
of roughly $5\arcdeg$ and a kinematic twist of about the same size
in the opposite direction.
Very approximately, the distortion to the velocity field
induced by a small triaxiality $T$ is $\delta V / V \sim T$, so
a $5\arcdeg$ kinematic twist might be characteristic of a weak
triaxiality $T\sim 0.1$. The photometric twist, if one assumes that
the true principal axes of the luminosity density surfaces are aligned,
signals a triaxiality {\em gradient\/}; but for small $T$, in order to observe
an isophotal twist of more than a degree or two requires a line of sight
for which the apparent ellipticity is small. Thus,
unless NGC 3379 is intrinsically twisted, the photometric and kinematic
data may well be indicating, completely independent of any arguments regarding
NGC 3115, a quite flattened, weakly triaxial system
seen in an orientation that makes it appear round.

\section{Conclusions\label{s.conclusions}}

We have measured the stellar kinematic profiles of NGC 3379 along four
position angles using the MMT. We have obtained mean velocities and dispersions
at excellent spatial resolution, with precisions better than $10\kms$
and frequently better than $5\kms$ out to $55\arcsec$, and at slightly lower
precision farther out. The $h_3$ and $h_4$ parameters are measured over
the entire slit length, and are found to be generally small. From a
Fourier reconstruction of the mean velocity field, we detect a $\sim
5\arcdeg$ twist of the kinematic major axis, over roughly the same range
of radii where the photometric major axis twists by $\sim 5\arcdeg$ in
the opposite direction. The most surprising aspect of our results
is the appearance of sharp features in the kinematic profiles.
There are sharp bends in the major-axis rotation curve, visible (though
less pronounced) on the diagonal position angles, which closely
resemble similar bends seen in the edge-on S0 NGC 3115.
Moreover, there is an abrupt flattening of the dispersion profile,
as well as local peaks in $h_3$ and $h_4$, all apparently
associated with the outer rotation curve bend near $17\arcsec$, and all
coinciding with a region where the surface photometry shows some of its
largest departures from an $r^{1/4}$ law.

The sharp kinematic transitions that we see in NGC 3379
are, as far as we know, unprecedented in any elliptical galaxy. But this is
much less a statement about galaxies than about data: no other elliptical
has been observed at this resolution over this large a range of radii.
The correspondence with kinematic features seen in NGC 3115
does not prove that NGC 3379 is an S0, since we do not know whether these
features are unique to S0's. Previously published data on NGC 3379 give the
impression of a gently rising rotation curve and a featureless, smoothly
falling dispersion profile. Except for the few systems identified
as having kinematically distinct cores, a cursory survey of the literature
gives a similar impression of most other ellipticals.

In the current standard conceptual picture, elliptical galaxies have smooth
and unremarkable rotation and dispersion profiles, except for a few
peculiar cases. Yet, one of the most ordinary ellipticals in the
sky, when examined with high enough precision, turns out to have far
richer dynamical structure
than expected. We should hardly be surprised to see this sort of thing
happen, though. Fifteen years ago it was also part of the standard
picture that elliptical galaxies had precisely elliptical isophotes,
except for a few peculiar cases. It was only after techniques had been
developed to measure departures from pure ellipses {\em at the 1\%
level\/} (Lauer \markcite{Lau85}1985, Jedrzejewski \markcite{Jed87}1987),
and after these measurements had been made for a respectable sample
of objects, that the distinction between the properties of
disky and boxy ellipticals (Bender et al.\ \markcite{Ben89}1989),
now regarded as fundamental, emerged. The potential of detailed
kinematic studies of elliptical galaxies to further elucidate their
structure and evolution remains, at this point, almost entirely unexplored.

\acknowledgments

TSS acknowledges support from NASA Astrophysical Theory Grant NAG5-3050 
and NSF CAREER grant AST-9703036.
We thank the director and staff of the Multiple
Mirror Telescope Observatory for their generous assistance and allocations
of time to this project. Ralf Bender kindly provided additional details
on his published data, and the anonymous referee helped us to improve the
paper by catching a number of errors.

\begin{deluxetable}{rrrrrrrrrrrrr}
	\scriptsize
\tablenum{1a}
\tablewidth{6.5in}
\tablecaption{Data for PA 70 (major axis)}
\tablehead{
\colhead{$R(\arcsec)$} &
\colhead{$V$} & \colhead{$\pm$} &
\colhead{$\sigma$} & \colhead{$\pm$} &
\colhead{$h_3$} & \colhead{$\pm$} &
\colhead{$h_4$} & \colhead{$\pm$} &
\colhead{Mean} & \colhead{$\pm$} &
\colhead{Disp.} & \colhead{$\pm$}}
\startdata
$-78.0$&$-26.0$&$15.8$&$183.1$&$17.8$&$0.153$&$0.133$&$-0.119$&$0.167$&$-23.9$&$14.6$&$183.4$&$22.4$\nl
$-55.2$&$-48.1$&$7.6$&$144.8$&$14.3$&$0.097$&$0.044$&$0.162$&$0.063$&$-41.7$&$7.3$&$149.0$&$10.4$\nl
$-41.4$&$-59.1$&$6.4$&$152.4$&$10.3$&$0.089$&$0.036$&$0.072$&$0.048$&$-53.6$&$6.0$&$154.0$&$8.6$\nl
$-33.6$&$-61.0$&$5.7$&$160.1$&$7.6$&$0.101$&$0.042$&$-0.003$&$0.055$&$-55.5$&$5.3$&$159.4$&$7.8$\nl
$-28.2$&$-51.4$&$5.1$&$160.6$&$6.5$&$0.060$&$0.038$&$-0.033$&$0.050$&$-48.0$&$5.0$&$160.5$&$7.4$\nl
$-24.0$&$-54.5$&$4.8$&$162.8$&$7.1$&$0.031$&$0.028$&$0.017$&$0.036$&$-52.7$&$4.8$&$162.4$&$7.0$\nl
$-21.0$&$-53.7$&$4.8$&$157.5$&$7.7$&$0.021$&$0.028$&$0.048$&$0.037$&$-52.1$&$4.7$&$159.3$&$6.8$\nl
$-18.6$&$-59.4$&$4.5$&$158.1$&$8.0$&$0.090$&$0.024$&$0.104$&$0.032$&$-53.2$&$4.2$&$160.7$&$6.1$\nl
$-16.8$&$-60.1$&$4.7$&$155.5$&$6.8$&$0.093$&$0.026$&$0.037$&$0.034$&$-54.1$&$4.2$&$156.1$&$6.1$\nl
$-15.6$&$-49.4$&$4.3$&$145.6$&$6.0$&$0.048$&$0.030$&$0.001$&$0.040$&$-46.0$&$4.0$&$144.0$&$5.9$\nl
$-14.4$&$-51.2$&$4.0$&$159.7$&$5.7$&$0.085$&$0.026$&$0.014$&$0.036$&$-46.5$&$3.7$&$156.2$&$5.5$\nl
$-13.2$&$-45.6$&$3.5$&$166.5$&$4.8$&$-0.022$&$0.020$&$-0.004$&$0.026$&$-46.9$&$3.5$&$166.4$&$5.1$\nl
$-12.0$&$-47.0$&$3.5$&$159.9$&$4.9$&$0.045$&$0.022$&$-0.001$&$0.030$&$-44.2$&$3.4$&$159.2$&$5.1$\nl
$-10.8$&$-47.4$&$3.0$&$172.4$&$4.0$&$0.030$&$0.017$&$-0.010$&$0.023$&$-45.7$&$3.0$&$172.0$&$4.5$\nl
$-9.6$&$-43.7$&$2.8$&$169.2$&$4.0$&$0.015$&$0.015$&$0.010$&$0.020$&$-42.9$&$2.8$&$169.3$&$4.1$\nl
$-8.4$&$-38.6$&$2.6$&$181.8$&$3.8$&$0.005$&$0.013$&$0.012$&$0.017$&$-38.4$&$2.7$&$182.0$&$3.9$\nl
$-7.2$&$-38.3$&$2.4$&$184.2$&$3.2$&$0.025$&$0.013$&$-0.014$&$0.017$&$-36.9$&$2.5$&$183.4$&$3.7$\nl
$-6.0$&$-35.1$&$2.2$&$188.8$&$2.8$&$0.023$&$0.012$&$-0.029$&$0.016$&$-33.9$&$2.3$&$188.0$&$3.5$\nl
$-4.8$&$-34.8$&$1.9$&$194.2$&$2.7$&$0.028$&$0.009$&$-0.007$&$0.012$&$-33.5$&$2.1$&$193.5$&$3.2$\nl
$-3.6$&$-34.3$&$1.8$&$196.1$&$2.5$&$0.041$&$0.009$&$-0.007$&$0.012$&$-25.7$&$3.8$&$193.3$&$6.1$\nl
$-2.4$&$-24.9$&$1.6$&$203.7$&$2.2$&$0.038$&$0.008$&$-0.012$&$0.010$&$-17.2$&$3.6$&$198.8$&$5.5$\nl
$-1.2$&$-13.4$&$1.5$&$214.1$&$2.1$&$0.031$&$0.007$&$-0.013$&$0.009$&$-7.7$&$3.7$&$208.1$&$5.5$\nl
$0.0$&$-1.4$&$1.5$&$218.1$&$2.1$&$0.015$&$0.006$&$-0.014$&$0.008$&$-1.4$&$3.7$&$211.2$&$5.5$\nl
$1.2$&$9.5$&$1.5$&$212.8$&$2.1$&$0.003$&$0.006$&$-0.016$&$0.008$&$5.7$&$3.6$&$205.6$&$5.2$\nl
$2.4$&$22.9$&$1.6$&$203.8$&$2.2$&$-0.009$&$0.007$&$-0.010$&$0.009$&$15.3$&$3.6$&$199.7$&$5.6$\nl
$3.6$&$32.5$&$1.7$&$199.4$&$2.4$&$-0.001$&$0.008$&$-0.006$&$0.010$&$27.3$&$3.9$&$196.4$&$6.4$\nl
$4.8$&$34.8$&$1.9$&$193.7$&$2.8$&$-0.007$&$0.009$&$0.011$&$0.012$&$34.5$&$2.1$&$193.7$&$3.1$\nl
$6.0$&$35.4$&$2.2$&$192.3$&$3.2$&$0.013$&$0.010$&$0.007$&$0.013$&$35.9$&$2.3$&$192.0$&$3.5$\nl
$7.2$&$38.5$&$2.5$&$188.0$&$3.3$&$-0.020$&$0.013$&$-0.012$&$0.016$&$37.7$&$2.6$&$188.4$&$3.8$\nl
$8.4$&$40.3$&$2.6$&$188.0$&$3.6$&$-0.018$&$0.013$&$-0.007$&$0.017$&$39.5$&$2.8$&$188.3$&$4.1$\nl
$9.6$&$45.3$&$2.8$&$177.1$&$3.7$&$-0.059$&$0.021$&$-0.037$&$0.027$&$42.9$&$2.9$&$178.1$&$4.3$\nl
$10.8$&$41.3$&$3.0$&$172.7$&$3.8$&$-0.036$&$0.019$&$-0.032$&$0.026$&$39.4$&$3.0$&$172.3$&$4.5$\nl
$12.0$&$47.8$&$3.2$&$164.6$&$4.7$&$-0.015$&$0.018$&$0.010$&$0.024$&$46.9$&$3.2$&$164.5$&$4.8$\nl
$13.2$&$47.0$&$3.6$&$166.0$&$5.4$&$-0.028$&$0.020$&$0.026$&$0.026$&$45.4$&$3.6$&$166.3$&$5.2$\nl
$14.4$&$59.6$&$3.9$&$168.2$&$5.8$&$-0.006$&$0.021$&$0.019$&$0.028$&$59.3$&$3.9$&$169.3$&$5.7$\nl
$15.6$&$56.8$&$4.2$&$165.7$&$5.7$&$-0.031$&$0.024$&$-0.003$&$0.032$&$54.9$&$4.1$&$165.0$&$6.0$\nl
$16.8$&$57.5$&$4.6$&$170.2$&$8.4$&$-0.003$&$0.024$&$0.076$&$0.031$&$58.5$&$4.7$&$168.0$&$6.7$\nl
$18.6$&$56.5$&$4.2$&$161.4$&$7.4$&$0.021$&$0.024$&$0.067$&$0.032$&$58.2$&$4.3$&$162.0$&$6.1$\nl
$21.0$&$60.0$&$4.9$&$154.8$&$6.8$&$-0.030$&$0.030$&$0.000$&$0.040$&$58.0$&$4.7$&$153.7$&$6.9$\nl
$24.0$&$56.2$&$4.8$&$168.3$&$6.1$&$-0.008$&$0.029$&$-0.028$&$0.041$&$55.7$&$4.8$&$167.9$&$6.9$\nl
$28.2$&$59.8$&$6.1$&$157.0$&$8.9$&$-0.153$&$0.079$&$-0.077$&$0.077$&$54.2$&$5.3$&$160.9$&$7.7$\nl
$33.6$&$53.9$&$5.4$&$165.2$&$8.2$&$0.004$&$0.030$&$0.028$&$0.039$&$53.8$&$5.5$&$165.8$&$7.9$\nl
$41.4$&$58.9$&$5.9$&$145.7$&$9.7$&$-0.026$&$0.037$&$0.056$&$0.050$&$56.7$&$5.9$&$149.2$&$8.5$\nl
$55.2$&$60.9$&$7.7$&$130.9$&$12.8$&$0.040$&$0.052$&$0.061$&$0.071$&$64.7$&$7.2$&$138.9$&$10.8$\nl
$78.0$&$19.4$&$16.1$&$103.1$&$21.5$&$0.209$&$0.132$&$0.289$&$0.196$&$43.5$&$13.4$&$137.4$&$20.6$\nl
\enddata
\end{deluxetable}

\begin{deluxetable}{rrrrrrrrrrrrr}
	\scriptsize
\tablenum{1b}
\tablewidth{6.5in}
\tablecaption{Data for PA 340 (minor axis)}
\tablehead{
\colhead{$R(\arcsec)$} &
\colhead{$V$} & \colhead{$\pm$} &
\colhead{$\sigma$} & \colhead{$\pm$} &
\colhead{$h_3$} & \colhead{$\pm$} &
\colhead{$h_4$} & \colhead{$\pm$} &
\colhead{Mean} & \colhead{$\pm$} &
\colhead{Disp.} & \colhead{$\pm$}}
\startdata
$-78.0$&$7.2$&$16.4$&$152.2$&$33.3$&$-0.056$&$0.095$&$0.133$&$0.131$&$6.1$&$15.3$&$142.5$&$22.6$\nl
$-55.2$&$8.6$&$10.0$&$115.5$&$14.0$&$0.058$&$0.085$&$0.007$&$0.116$&$13.3$&$8.1$&$117.3$&$12.7$\nl
$-41.4$&$3.4$&$6.6$&$135.8$&$11.1$&$0.023$&$0.044$&$0.065$&$0.059$&$5.6$&$6.3$&$140.6$&$9.1$\nl
$-33.6$&$-9.4$&$5.7$&$157.9$&$7.8$&$-0.001$&$0.034$&$-0.004$&$0.044$&$-9.5$&$5.7$&$157.8$&$8.3$\nl
$-28.2$&$-6.8$&$5.8$&$166.0$&$6.7$&$0.011$&$0.049$&$-0.090$&$0.081$&$-6.5$&$5.7$&$163.5$&$8.4$\nl
$-24.0$&$-8.7$&$5.3$&$155.7$&$6.3$&$-0.134$&$0.064$&$-0.158$&$0.095$&$-10.4$&$5.0$&$159.7$&$7.4$\nl
$-21.0$&$4.1$&$5.3$&$152.3$&$6.4$&$-0.037$&$0.064$&$-0.147$&$0.108$&$2.7$&$5.2$&$151.0$&$7.7$\nl
$-18.6$&$7.2$&$4.5$&$180.6$&$5.7$&$0.025$&$0.027$&$-0.032$&$0.038$&$8.5$&$4.5$&$179.3$&$6.5$\nl
$-16.8$&$7.5$&$4.4$&$171.6$&$6.3$&$-0.005$&$0.024$&$0.009$&$0.031$&$7.1$&$4.4$&$172.0$&$6.4$\nl
$-15.6$&$3.8$&$4.2$&$158.1$&$5.3$&$-0.035$&$0.031$&$-0.032$&$0.043$&$2.1$&$4.0$&$158.1$&$5.9$\nl
$-14.4$&$1.2$&$4.2$&$155.1$&$5.7$&$0.015$&$0.026$&$-0.007$&$0.034$&$2.2$&$4.0$&$154.3$&$5.9$\nl
$-13.2$&$2.4$&$3.6$&$165.8$&$5.4$&$0.019$&$0.020$&$0.020$&$0.027$&$3.5$&$3.7$&$166.0$&$5.3$\nl
$-12.0$&$2.3$&$3.3$&$168.7$&$4.7$&$0.047$&$0.018$&$0.013$&$0.024$&$5.0$&$3.3$&$168.6$&$4.8$\nl
$-10.8$&$2.1$&$3.0$&$168.8$&$4.1$&$0.063$&$0.018$&$0.000$&$0.024$&$5.6$&$2.9$&$168.6$&$4.3$\nl
$-9.6$&$0.8$&$2.7$&$172.5$&$3.8$&$0.034$&$0.014$&$0.006$&$0.019$&$2.7$&$2.7$&$172.4$&$4.0$\nl
$-8.4$&$-0.1$&$2.6$&$178.3$&$3.7$&$0.032$&$0.013$&$0.007$&$0.017$&$1.5$&$2.7$&$177.8$&$3.9$\nl
$-7.2$&$-0.4$&$2.3$&$185.9$&$3.3$&$0.024$&$0.012$&$-0.004$&$0.015$&$0.8$&$2.5$&$185.4$&$3.7$\nl
$-6.0$&$2.1$&$2.2$&$189.4$&$3.0$&$0.014$&$0.011$&$-0.008$&$0.014$&$2.8$&$2.3$&$189.0$&$3.4$\nl
$-4.8$&$4.2$&$1.9$&$191.1$&$2.6$&$0.022$&$0.010$&$-0.011$&$0.013$&$5.2$&$2.1$&$190.6$&$3.2$\nl
$-3.6$&$2.8$&$1.7$&$193.9$&$2.4$&$0.016$&$0.008$&$-0.004$&$0.011$&$2.1$&$3.9$&$191.9$&$6.7$\nl
$-2.4$&$1.2$&$1.5$&$206.9$&$2.2$&$0.020$&$0.007$&$-0.002$&$0.009$&$1.7$&$3.9$&$206.0$&$6.5$\nl
$-1.2$&$0.7$&$1.4$&$208.7$&$2.0$&$0.014$&$0.006$&$-0.014$&$0.008$&$-0.7$&$3.5$&$202.0$&$5.2$\nl
$0.0$&$-1.9$&$1.4$&$210.8$&$1.9$&$0.018$&$0.006$&$-0.020$&$0.008$&$-1.9$&$3.3$&$201.9$&$4.5$\nl
$1.2$&$-1.6$&$1.4$&$212.8$&$1.9$&$0.023$&$0.006$&$-0.015$&$0.008$&$-0.0$&$3.5$&$205.8$&$5.2$\nl
$2.4$&$-1.8$&$1.5$&$205.0$&$2.0$&$0.014$&$0.007$&$-0.016$&$0.009$&$-3.3$&$3.5$&$197.7$&$5.1$\nl
$3.6$&$-1.1$&$1.7$&$195.4$&$2.4$&$0.028$&$0.008$&$-0.004$&$0.011$&$2.2$&$3.9$&$193.7$&$6.7$\nl
$4.8$&$0.0$&$1.9$&$192.3$&$2.6$&$0.014$&$0.009$&$-0.007$&$0.012$&$0.6$&$2.0$&$192.2$&$3.1$\nl
$6.0$&$-4.5$&$2.1$&$184.2$&$2.9$&$-0.005$&$0.010$&$0.000$&$0.014$&$-4.8$&$2.2$&$184.3$&$3.3$\nl
$7.2$&$-5.5$&$2.3$&$178.4$&$3.1$&$-0.009$&$0.012$&$-0.018$&$0.017$&$-6.0$&$2.4$&$177.9$&$3.6$\nl
$8.4$&$-1.3$&$2.6$&$172.1$&$3.7$&$-0.013$&$0.014$&$0.004$&$0.018$&$-2.0$&$2.7$&$172.2$&$4.0$\nl
$9.6$&$2.9$&$2.8$&$173.6$&$4.0$&$-0.022$&$0.015$&$0.007$&$0.019$&$1.6$&$2.9$&$174.1$&$4.2$\nl
$10.8$&$-5.5$&$3.0$&$170.2$&$3.8$&$0.038$&$0.020$&$-0.035$&$0.027$&$-3.5$&$3.0$&$168.8$&$4.5$\nl
$12.0$&$-8.8$&$3.2$&$163.5$&$4.8$&$0.020$&$0.018$&$0.020$&$0.024$&$-7.5$&$3.3$&$164.1$&$4.7$\nl
$13.2$&$-0.5$&$3.6$&$163.1$&$5.2$&$-0.012$&$0.021$&$0.013$&$0.027$&$-1.2$&$3.6$&$164.0$&$5.3$\nl
$14.4$&$-2.3$&$3.8$&$163.8$&$5.7$&$0.008$&$0.021$&$0.022$&$0.028$&$-1.7$&$3.8$&$164.9$&$5.5$\nl
$15.6$&$-2.4$&$4.2$&$163.3$&$5.8$&$0.015$&$0.024$&$-0.002$&$0.031$&$-1.5$&$4.1$&$163.1$&$6.0$\nl
$16.8$&$0.7$&$4.8$&$153.1$&$6.5$&$-0.013$&$0.029$&$-0.004$&$0.038$&$-0.1$&$4.6$&$153.1$&$6.7$\nl
$18.6$&$-0.7$&$4.5$&$148.3$&$7.9$&$0.061$&$0.027$&$0.080$&$0.037$&$4.0$&$4.4$&$156.2$&$6.2$\nl
$21.0$&$-3.2$&$4.9$&$145.9$&$6.9$&$0.082$&$0.035$&$0.014$&$0.048$&$2.0$&$4.5$&$147.7$&$6.6$\nl
$24.0$&$-2.6$&$5.0$&$152.9$&$7.8$&$0.056$&$0.029$&$0.046$&$0.039$&$1.0$&$4.8$&$156.0$&$7.0$\nl
$28.2$&$-3.7$&$5.5$&$162.6$&$6.4$&$0.113$&$0.054$&$-0.088$&$0.067$&$-0.1$&$5.2$&$163.5$&$7.6$\nl
$33.6$&$1.2$&$6.0$&$166.8$&$6.8$&$-0.000$&$0.051$&$-0.097$&$0.086$&$2.5$&$6.1$&$163.9$&$9.0$\nl
$41.4$&$10.4$&$7.2$&$160.9$&$9.1$&$-0.009$&$0.075$&$-0.109$&$0.130$&$9.7$&$7.2$&$158.9$&$10.7$\nl
$55.2$&$27.5$&$10.1$&$150.9$&$10.8$&$-0.093$&$0.124$&$-0.181$&$0.185$&$25.2$&$9.6$&$147.6$&$15.0$\nl
$78.0$&$-4.2$&$17.9$&$126.7$&$23.1$&$-0.047$&$0.464$&$-0.617$&$0.819$&$-6.1$&$19.2$&$133.7$&$29.5$\nl
\enddata
\end{deluxetable}

\begin{deluxetable}{rrrrrrrrrrrrr}
	\scriptsize
\tablenum{1c}
\tablewidth{6.5in}
\tablecaption{Data for PA 25}
\tablehead{
\colhead{$R(\arcsec)$} &
\colhead{$V$} & \colhead{$\pm$} &
\colhead{$\sigma$} & \colhead{$\pm$} &
\colhead{$h_3$} & \colhead{$\pm$} &
\colhead{$h_4$} & \colhead{$\pm$} &
\colhead{Mean} & \colhead{$\pm$} &
\colhead{Disp.} & \colhead{$\pm$}}
\startdata
$-78.0$&$-16.7$&$14.7$&$170.2$&$17.4$&$0.150$&$0.157$&$-0.177$&$0.229$&$-18.0$&$14.0$&$180.6$&$21.6$\nl
$-55.2$&$-17.3$&$7.7$&$133.5$&$11.1$&$0.082$&$0.056$&$0.020$&$0.078$&$-11.4$&$6.8$&$134.3$&$10.4$\nl
$-41.4$&$-27.4$&$5.2$&$133.9$&$8.2$&$-0.008$&$0.035$&$0.044$&$0.047$&$-27.5$&$5.0$&$134.8$&$7.3$\nl
$-33.6$&$-34.1$&$5.7$&$153.9$&$6.1$&$0.188$&$0.061$&$-0.127$&$0.071$&$-29.1$&$4.9$&$159.5$&$7.5$\nl
$-28.2$&$-34.1$&$4.8$&$160.1$&$9.6$&$0.070$&$0.026$&$0.142$&$0.035$&$-28.1$&$4.8$&$163.8$&$6.8$\nl
$-24.0$&$-31.9$&$4.7$&$175.1$&$7.9$&$0.067$&$0.024$&$0.065$&$0.030$&$-27.0$&$4.8$&$180.6$&$6.7$\nl
$-21.0$&$-31.4$&$4.7$&$164.4$&$8.0$&$0.004$&$0.026$&$0.058$&$0.034$&$-31.5$&$4.7$&$165.6$&$6.8$\nl
$-18.6$&$-28.4$&$4.0$&$155.0$&$6.6$&$-0.002$&$0.023$&$0.053$&$0.031$&$-28.7$&$3.9$&$154.8$&$5.6$\nl
$-16.8$&$-27.6$&$4.3$&$164.2$&$6.1$&$-0.017$&$0.024$&$0.009$&$0.032$&$-28.6$&$4.3$&$164.7$&$6.1$\nl
$-15.6$&$-31.7$&$4.0$&$175.2$&$6.7$&$0.011$&$0.021$&$0.048$&$0.027$&$-30.9$&$4.2$&$175.6$&$5.9$\nl
$-14.4$&$-27.8$&$3.7$&$164.8$&$6.4$&$-0.004$&$0.020$&$0.064$&$0.027$&$-28.3$&$3.8$&$167.5$&$5.4$\nl
$-13.2$&$-23.4$&$3.3$&$173.0$&$5.2$&$-0.006$&$0.017$&$0.034$&$0.023$&$-23.9$&$3.4$&$174.2$&$4.9$\nl
$-12.0$&$-26.8$&$3.0$&$166.0$&$5.2$&$0.018$&$0.017$&$0.059$&$0.022$&$-25.7$&$3.2$&$169.5$&$4.5$\nl
$-10.8$&$-27.9$&$2.9$&$173.1$&$4.4$&$0.020$&$0.016$&$0.024$&$0.020$&$-26.8$&$3.0$&$173.7$&$4.4$\nl
$-9.6$&$-28.1$&$2.7$&$170.7$&$3.7$&$0.032$&$0.015$&$0.003$&$0.019$&$-26.2$&$2.7$&$170.9$&$4.0$\nl
$-8.4$&$-25.7$&$2.5$&$173.2$&$3.7$&$0.041$&$0.013$&$0.020$&$0.017$&$-23.2$&$2.5$&$174.0$&$3.8$\nl
$-7.2$&$-27.8$&$2.2$&$183.2$&$3.1$&$0.001$&$0.011$&$-0.000$&$0.015$&$-27.7$&$2.4$&$183.2$&$3.5$\nl
$-6.0$&$-25.0$&$2.0$&$193.8$&$2.8$&$0.028$&$0.009$&$0.002$&$0.012$&$-23.8$&$2.1$&$193.6$&$3.2$\nl
$-4.8$&$-25.9$&$1.9$&$189.8$&$2.6$&$0.033$&$0.009$&$-0.001$&$0.012$&$-24.4$&$2.0$&$189.6$&$3.0$\nl
$-3.6$&$-20.3$&$1.7$&$198.8$&$2.2$&$0.023$&$0.008$&$-0.022$&$0.011$&$-19.7$&$3.4$&$189.6$&$4.8$\nl
$-2.4$&$-16.6$&$1.5$&$205.5$&$2.1$&$0.027$&$0.007$&$-0.011$&$0.009$&$-14.5$&$3.7$&$200.2$&$5.8$\nl
$-1.2$&$-9.9$&$1.4$&$209.1$&$2.0$&$0.029$&$0.006$&$-0.012$&$0.008$&$-6.9$&$3.5$&$203.5$&$5.5$\nl
$0.0$&$-1.5$&$1.4$&$209.4$&$2.0$&$0.020$&$0.006$&$-0.012$&$0.008$&$-1.5$&$3.6$&$203.5$&$5.5$\nl
$1.2$&$6.4$&$1.4$&$209.1$&$1.9$&$0.007$&$0.006$&$-0.021$&$0.008$&$2.3$&$3.3$&$200.2$&$4.5$\nl
$2.4$&$11.4$&$1.5$&$204.8$&$2.1$&$0.002$&$0.007$&$-0.004$&$0.009$&$4.9$&$3.7$&$202.8$&$6.0$\nl
$3.6$&$16.5$&$1.6$&$198.4$&$2.4$&$0.007$&$0.008$&$0.005$&$0.010$&$11.9$&$3.9$&$200.8$&$6.8$\nl
$4.8$&$19.5$&$1.8$&$197.6$&$2.6$&$-0.007$&$0.009$&$-0.009$&$0.011$&$19.3$&$2.0$&$197.9$&$3.1$\nl
$6.0$&$20.3$&$2.2$&$192.3$&$3.0$&$-0.012$&$0.010$&$-0.005$&$0.014$&$19.8$&$2.3$&$192.4$&$3.5$\nl
$7.2$&$21.8$&$2.3$&$183.7$&$3.3$&$-0.007$&$0.012$&$-0.002$&$0.015$&$21.4$&$2.5$&$183.7$&$3.7$\nl
$8.4$&$24.9$&$2.5$&$174.8$&$3.4$&$-0.026$&$0.014$&$-0.006$&$0.018$&$23.4$&$2.6$&$174.6$&$3.8$\nl
$9.6$&$19.7$&$2.7$&$177.2$&$3.6$&$-0.025$&$0.015$&$-0.016$&$0.020$&$18.4$&$2.8$&$176.8$&$4.1$\nl
$10.8$&$28.2$&$2.7$&$175.3$&$3.8$&$-0.005$&$0.015$&$-0.001$&$0.019$&$27.9$&$2.8$&$175.2$&$4.2$\nl
$12.0$&$28.3$&$3.1$&$164.8$&$4.4$&$0.011$&$0.017$&$0.009$&$0.022$&$29.1$&$3.1$&$165.2$&$4.5$\nl
$13.2$&$26.0$&$3.5$&$171.5$&$4.9$&$0.024$&$0.019$&$0.009$&$0.024$&$27.3$&$3.5$&$171.7$&$5.1$\nl
$14.4$&$30.9$&$3.6$&$167.7$&$5.6$&$0.023$&$0.020$&$0.034$&$0.026$&$32.1$&$3.7$&$168.5$&$5.3$\nl
$15.6$&$33.3$&$4.2$&$134.3$&$8.1$&$-0.044$&$0.028$&$0.110$&$0.040$&$29.8$&$3.9$&$140.5$&$5.6$\nl
$16.8$&$32.6$&$4.6$&$163.8$&$7.7$&$-0.018$&$0.026$&$0.054$&$0.034$&$31.3$&$4.6$&$166.8$&$6.6$\nl
$18.6$&$30.4$&$4.0$&$164.0$&$6.8$&$-0.054$&$0.022$&$0.065$&$0.029$&$25.9$&$4.1$&$170.1$&$5.8$\nl
$21.0$&$34.8$&$4.4$&$176.8$&$6.5$&$0.011$&$0.023$&$0.016$&$0.030$&$35.4$&$4.6$&$177.4$&$6.5$\nl
$24.0$&$34.3$&$4.6$&$153.2$&$6.8$&$-0.020$&$0.028$&$0.019$&$0.037$&$32.9$&$4.6$&$154.0$&$6.6$\nl
$28.2$&$40.1$&$5.1$&$148.9$&$6.3$&$-0.058$&$0.043$&$-0.038$&$0.057$&$37.2$&$4.8$&$148.7$&$7.1$\nl
$33.6$&$40.5$&$5.3$&$144.0$&$9.0$&$-0.065$&$0.032$&$0.076$&$0.044$&$35.4$&$5.1$&$148.1$&$7.2$\nl
$41.4$&$31.3$&$5.5$&$152.3$&$8.4$&$-0.059$&$0.032$&$0.041$&$0.043$&$27.5$&$5.4$&$152.8$&$7.9$\nl
$55.2$&$36.9$&$7.7$&$113.5$&$14.1$&$-0.055$&$0.058$&$0.101$&$0.086$&$33.0$&$6.5$&$119.0$&$9.8$\nl
$78.0$&$37.0$&$14.4$&$178.0$&$15.8$&$-0.103$&$0.142$&$-0.214$&$0.230$&$38.5$&$14.8$&$183.7$&$22.6$\nl
\enddata
\end{deluxetable}

\begin{deluxetable}{rrrrrrrrrrrrr}
	\scriptsize
\tablenum{1d}
\tablewidth{6.5in}
\tablecaption{Data for PA 115 }
\tablehead{
\colhead{$R(\arcsec)$} &
\colhead{$V$} & \colhead{$\pm$} &
\colhead{$\sigma$} & \colhead{$\pm$} &
\colhead{$h_3$} & \colhead{$\pm$} &
\colhead{$h_4$} & \colhead{$\pm$} &
\colhead{Mean} & \colhead{$\pm$} &
\colhead{Disp.} & \colhead{$\pm$}}
\startdata
$-78.0$&$-62.0$&$19.3$&$214.4$&$23.0$&$-0.042$&$0.099$&$-0.060$&$0.143$&$-63.7$&$19.2$&$208.9$&$28.3$\nl
$-55.2$&$-42.0$&$8.9$&$134.2$&$14.5$&$0.152$&$0.056$&$0.199$&$0.079$&$-30.2$&$8.1$&$147.1$&$11.8$\nl
$-41.4$&$-48.1$&$6.3$&$149.8$&$9.8$&$0.078$&$0.037$&$0.055$&$0.049$&$-43.9$&$5.9$&$150.4$&$8.6$\nl
$-33.6$&$-33.9$&$5.8$&$152.4$&$7.7$&$0.025$&$0.039$&$-0.015$&$0.052$&$-32.1$&$5.5$&$150.5$&$8.2$\nl
$-28.2$&$-33.7$&$5.0$&$164.4$&$7.4$&$-0.007$&$0.028$&$0.015$&$0.037$&$-34.1$&$5.1$&$164.6$&$7.4$\nl
$-24.0$&$-36.1$&$5.4$&$167.6$&$7.1$&$0.092$&$0.042$&$-0.017$&$0.052$&$-31.5$&$5.1$&$167.6$&$7.4$\nl
$-21.0$&$-39.0$&$5.4$&$167.2$&$8.6$&$0.123$&$0.027$&$0.089$&$0.034$&$-30.3$&$4.9$&$173.1$&$7.1$\nl
$-18.6$&$-31.4$&$4.6$&$165.3$&$6.2$&$0.142$&$0.052$&$-0.092$&$0.056$&$-27.2$&$4.2$&$169.7$&$6.1$\nl
$-16.8$&$-39.3$&$5.2$&$161.7$&$8.0$&$0.075$&$0.028$&$0.046$&$0.037$&$-34.7$&$4.9$&$163.6$&$7.0$\nl
$-15.6$&$-37.6$&$4.4$&$169.7$&$6.6$&$0.065$&$0.023$&$0.031$&$0.030$&$-34.0$&$4.3$&$171.7$&$6.2$\nl
$-14.4$&$-32.5$&$4.0$&$157.2$&$6.5$&$0.042$&$0.023$&$0.054$&$0.030$&$-29.2$&$3.9$&$160.8$&$5.6$\nl
$-13.2$&$-35.0$&$3.6$&$163.7$&$5.5$&$0.015$&$0.020$&$0.025$&$0.027$&$-34.0$&$3.6$&$164.5$&$5.3$\nl
$-12.0$&$-35.7$&$3.5$&$160.1$&$5.3$&$0.056$&$0.019$&$0.041$&$0.025$&$-31.7$&$3.4$&$163.3$&$4.9$\nl
$-10.8$&$-31.3$&$3.2$&$169.7$&$4.5$&$0.057$&$0.018$&$0.008$&$0.024$&$-27.8$&$3.1$&$170.7$&$4.6$\nl
$-9.6$&$-32.6$&$2.9$&$169.5$&$3.9$&$0.038$&$0.017$&$-0.006$&$0.022$&$-30.3$&$2.9$&$168.7$&$4.2$\nl
$-8.4$&$-29.2$&$2.6$&$176.0$&$3.9$&$0.007$&$0.014$&$0.010$&$0.018$&$-28.9$&$2.7$&$175.9$&$4.0$\nl
$-7.2$&$-28.5$&$2.5$&$189.2$&$3.5$&$0.017$&$0.012$&$-0.008$&$0.016$&$-27.7$&$2.6$&$189.2$&$3.9$\nl
$-6.0$&$-27.2$&$2.2$&$190.1$&$3.2$&$0.026$&$0.011$&$0.009$&$0.014$&$-26.2$&$2.3$&$189.6$&$3.5$\nl
$-4.8$&$-24.1$&$2.0$&$192.9$&$2.9$&$0.028$&$0.009$&$0.003$&$0.012$&$-22.9$&$2.1$&$192.7$&$3.2$\nl
$-3.6$&$-22.2$&$1.8$&$197.4$&$2.5$&$0.022$&$0.008$&$-0.008$&$0.011$&$-20.3$&$4.0$&$193.7$&$6.7$\nl
$-2.4$&$-19.3$&$1.6$&$203.7$&$2.3$&$0.031$&$0.007$&$-0.001$&$0.009$&$-13.8$&$3.9$&$203.1$&$6.4$\nl
$-1.2$&$-10.5$&$1.5$&$210.2$&$2.1$&$0.024$&$0.007$&$-0.009$&$0.009$&$-7.6$&$3.8$&$206.0$&$6.0$\nl
$0.0$&$3.3$&$1.5$&$211.6$&$2.1$&$0.016$&$0.006$&$-0.009$&$0.008$&$3.3$&$3.8$&$207.1$&$6.0$\nl
$1.2$&$9.0$&$1.5$&$204.4$&$2.0$&$0.004$&$0.007$&$-0.016$&$0.009$&$5.3$&$3.4$&$197.4$&$5.0$\nl
$2.4$&$17.9$&$1.6$&$200.7$&$2.2$&$-0.003$&$0.008$&$-0.014$&$0.010$&$11.8$&$3.6$&$194.8$&$5.4$\nl
$3.6$&$15.9$&$1.7$&$196.6$&$2.4$&$-0.002$&$0.008$&$-0.009$&$0.011$&$10.2$&$3.8$&$192.7$&$6.1$\nl
$4.8$&$24.8$&$1.9$&$192.3$&$2.7$&$0.007$&$0.009$&$-0.007$&$0.012$&$25.2$&$2.1$&$192.2$&$3.2$\nl
$6.0$&$25.6$&$2.2$&$189.8$&$3.2$&$-0.013$&$0.011$&$0.005$&$0.014$&$25.0$&$2.3$&$189.9$&$3.5$\nl
$7.2$&$29.3$&$2.6$&$186.1$&$3.6$&$-0.022$&$0.013$&$-0.003$&$0.017$&$28.3$&$2.7$&$186.2$&$4.0$\nl
$8.4$&$28.3$&$2.6$&$179.8$&$3.4$&$0.016$&$0.015$&$-0.026$&$0.020$&$29.1$&$2.7$&$179.2$&$4.0$\nl
$9.6$&$32.5$&$2.8$&$169.9$&$4.2$&$-0.008$&$0.015$&$0.026$&$0.019$&$31.9$&$2.9$&$170.9$&$4.2$\nl
$10.8$&$30.5$&$3.3$&$169.7$&$5.0$&$-0.027$&$0.018$&$0.030$&$0.023$&$28.8$&$3.3$&$170.8$&$4.8$\nl
$12.0$&$27.5$&$3.6$&$167.9$&$5.8$&$0.017$&$0.020$&$0.042$&$0.026$&$28.4$&$3.7$&$169.2$&$5.3$\nl
$13.2$&$32.7$&$3.7$&$159.6$&$5.8$&$-0.017$&$0.022$&$0.034$&$0.028$&$31.9$&$3.7$&$161.5$&$5.4$\nl
$14.4$&$31.6$&$4.0$&$142.3$&$6.9$&$-0.034$&$0.025$&$0.075$&$0.034$&$29.1$&$3.7$&$145.0$&$5.4$\nl
$15.6$&$40.8$&$4.5$&$147.8$&$7.5$&$-0.026$&$0.027$&$0.060$&$0.037$&$39.3$&$4.3$&$151.0$&$6.2$\nl
$16.8$&$42.3$&$5.2$&$141.6$&$10.1$&$-0.045$&$0.032$&$0.122$&$0.045$&$39.3$&$4.8$&$146.4$&$7.0$\nl
$18.6$&$35.2$&$4.3$&$160.1$&$6.9$&$-0.026$&$0.024$&$0.047$&$0.032$&$33.4$&$4.3$&$162.9$&$6.2$\nl
$21.0$&$32.1$&$4.9$&$151.4$&$7.5$&$-0.068$&$0.029$&$0.040$&$0.039$&$27.3$&$4.7$&$155.2$&$6.7$\nl
$24.0$&$41.2$&$5.3$&$153.9$&$6.6$&$-0.096$&$0.045$&$-0.030$&$0.057$&$36.0$&$4.9$&$154.1$&$7.2$\nl
$28.2$&$37.8$&$5.4$&$159.1$&$8.0$&$-0.061$&$0.030$&$0.034$&$0.040$&$33.7$&$5.3$&$162.5$&$7.5$\nl
$33.6$&$42.4$&$5.6$&$143.0$&$7.9$&$0.038$&$0.036$&$0.009$&$0.048$&$44.9$&$5.3$&$142.8$&$7.8$\nl
$41.4$&$36.2$&$6.0$&$143.0$&$12.2$&$0.014$&$0.037$&$0.123$&$0.053$&$36.6$&$6.1$&$151.8$&$8.6$\nl
$55.2$&$38.1$&$8.5$&$134.1$&$13.6$&$-0.068$&$0.056$&$0.057$&$0.075$&$32.9$&$7.7$&$137.8$&$11.6$\nl
$78.0$&$35.6$&$36.7$&$72.2$&$44.2$&$-0.241$&$0.721$&$0.048$&$1.027$&$22.6$&$10.4$&$66.7$&$18.1$\nl
\enddata
\end{deluxetable}

\end{document}